\documentclass[12pt,a4paper,oneside]{article}
\usepackage{jheppub}
\usepackage{graphicx,amssymb,amsmath,color}
\usepackage{cancel}
\usepackage{placeins}

\def\beq{\begin{equation}}
\def\eeq{\end{equation}}
\def\bea{\begin{eqnarray}}
\def\eea{\end{eqnarray}}

\def\bit{\begin{itemize}}
\def\eit{\end{itemize}}

\def\baa{\begin{array}}
\def\eaa{\end{array}}

\def\simgt{\mathrel{\lower2.5pt\vbox{\lineskip=0pt\baselineskip=0pt
           \hbox{$>$}\hbox{$\sim$}}}}
\def\simlt{\mathrel{\lower2.5pt\vbox{\lineskip=0pt\baselineskip=0pt
           \hbox{$<$}\hbox{$\sim$}}}}

\def\bfc{\begin{figure}\begin{center}}
\def\efc{\end{center}\end{figure}}

\def\bal#1\eal{\begin{align}#1\end{align}}

\def\lsim{\lesssim}

\def\GeV{{\rm GeV}}
\def\TeV{{\rm TeV}}

\def\PeV{{\rm PeV}}

\definecolor{chromeyellow}{rgb}{1.0, 0.65, 0.0}
\definecolor{darkcoral}{rgb}{0.8, 0.36, 0.27}
\definecolor{cadmiumgreen}{rgb}{0.0, 0.42, 0.24}

\title{What happens when supercooling is terminated 
by curvature flipping of the effective potential?}

\author[a]{Tomasz P. Dutka}
\author[b]{Tae Hyun Jung}
\author[a,b,c]{Chang Sub Shin}

\affiliation[a]{School of Physics, Korea Institute for Advanced Study, Seoul, 02455, Republic of Korea}
\affiliation[b]{Particle Theory  and Cosmology Group, Center for Theoretical Physics of the Universe,
Institute for Basic Science (IBS),
 Daejeon, 34126, Korea}
\affiliation[c]{Department of Physics and Institute of Quantum Systems, 
Chungnam National University, Daejeon 34134, Korea}

\emailAdd{tdutka@kias.re.kr}
\emailAdd{thjung0720@gmail.com}
\emailAdd{csshin@cnu.ac.kr}

\abstract{
We explore the nature of a certain type of supercooled phase transition, 
where the supercooling is guaranteed to end due to the curvature of the finite-temperature effective potential at the origin experiencing a sign flip at some temperature. 
In such models the potential barrier trapping the scalar field at the meta-stable origin is quickly vanishing at the temperature scale of the phase transition.
It is therefore not immediately clear if critical bubbles are able to form, or whether the field will simply transition over the barrier and smoothly roll down to the true minimum.
To address this question, we perform lattice simulations of a scalar potential exhibiting supercooling, with a small barrier around the origin, and qualitatively determine the fate of the phase transition. 
Our simulations indicate that, owing to the required flatness of the potential, the scalar field remains trapped around the origin such that the phase transition generically proceeds via the nucleation and expansion of true-vacuum bubbles. 
We comment on the possible gravitational wave signals one might expect in a concrete toy model and discuss the parameter space in which bubble percolation is and isn't expected. 
}

\preprint{CTPU-PTC-24-40}

\arxivnumber{2412.15864}

\begin{document}
\maketitle
\flushbottom
\newpage
\section{Introduction}
Cosmological first-order phase transition is a hypothetical phenomenon motivated in various contexts.
In the early Universe, a symmetry broken spontaneously is often restored at high temperatures, and when the temperature drops, due to the expansion of the Universe, a first-order phase transition may occur.
Although the standard model (SM) of particle physics predicts no cosmic first-order phase transition, there are still unsolved mysteries in the SM which necessitate new physics, and can source such a phase transition.
For instance, many theoretical models beyond the SM include modifications to the SM Higgs sector, making the electroweak phase transition itself first-order, or new symmetries can exist at higher scales that exhibit a first-order phase transition.
Additionally, the dynamics of the first-order phase transition itself can also play a crucial role in currently unsolved issues within the SM.
For example, the dynamics of the expanding bubbles have been shown to have important, non-negligible consequences on the number density of dark matter or the baryon asymmetry~\cite{
Chway:2019kft, Baker:2019ndr, Ahmadvand:2021vxs, Hong:2020est, Azatov:2021ifm, Asadi:2021pwo,
Shaposhnikov:1987tw, Cline:2018fuq, Hall:2019ank, Fujikura:2021abj, Baldes:2021vyz, Azatov:2021irb, Arakawa:2021wgz, Huang:2022vkf, Dasgupta:2022isg, Chun:2023ezg}, 
and it can provide a primordial black hole (PBH) formation mechanisms alternative to the typical inflationary scenarios\,\cite{
Hawking:1982ga, Moss:1994pi, Jung:2021mku,
Sato:1981bf, Maeda:1981gw, Kodama:1982sf, Hall:1989hr, Kusenko:2020pcg, 
Khlopov:1999ys, Lewicki:2019gmv, 
Gross:2021qgx, Baker:2021nyl, Kawana:2021tde, Baker:2021sno,
Liu:2021svg, Hashino:2021qoq, Huang:2022him, DeLuca:2022bjs, Kawana:2022olo, Lewicki:2023ioy, Gouttenoire:2023naa, Jinno:2023vnr, Ai:2024cka}.
A strongly supercooled phase transition can also provide a dilution of any dangerous relics which are predicted in many motivated models\,\cite{Lyth:1995hj, Lyth:1995ka, Barreiro:1996dx, Jeong:2004hy, Easther:2008sx}.
As first-order phase transition generically produces sufficiently strong stochastic gravitational wave signals\,\cite{Witten:1984rs,Hogan:1986qda,Kosowsky:1991ua,Kosowsky:1992vn,Kamionkowski:1993fg,Espinosa:2010hh,Caprini:2019egz, Schmitz:2020syl}, and the aforementioned mechanisms typically require a sufficiently strong transition (with supercooling), there are exciting prospects for these models to be tested and observed in future gravitational wave detectors.

A strongly supercooled phase transition requires the symmetry-breaking potential to be effectively \emph{flat}.
Here, the flatness is defined as 
$\Delta V/ v_\phi^4 \ll 1$ where $\Delta V$ is the vacuum energy difference between the symmetry-breaking phase and symmetry-restored phase, and $v_\phi$ is associated with the symmetry-breaking scale (typically the vacuum expectation value of the symmetry-breaking scalar field).
For instance, such a flat potential can be provided by a classically scale-invariant potential with radiative breaking effects, in which the potential has the form of $\kappa_4 \phi^4 \log (\phi/\mu_*)$, where $\mu_*$ is the renormalization scale at which the tree-level potential vanishes.
These class of potentials are known to have a long supercooling period, resulting in a large dilution factor, and a run-away (or ultra-relativistic) bubble wall, however $\kappa_4$ is restricted to be greater than a certain value (depending on $\mu_*$) in order to avoid eternal inflation, see e.g.~\cite{Levi:2022bzt,Marzo:2018nov,Lewicki:2021xku}.
Since the coefficient, $\kappa_4$, is given by the one-loop Coleman-Weinberg correction~\cite{PhysRevD.7.1888}, its size is also related to the interaction strength of the scalar field within the plasma.
These relations put limitations on the model's applicability in phenomenological studies.

Another example of a flat potential can be found in supersymmetric (SUSY) theories.
With SUSY, scalar quartic couplings arise from a superpotential and can even be absent.
For instance, one can consider the potential for $\phi$ (typically referred to as a ``flaton") in which its self-quartic coupling vanishes, the curvature at the origin is negative, and stabilization of the potential is generated through loop effects.
In this case, the flaton potential scales as $V(\phi) \sim m^2 \phi^2 \log (\phi/\mu_*)$, which is even flatter than classically scale-invariant models, so it drives a strong period of supercooling which starts at the critical temperature, $T_c$.
Contrary to scale-invariant models, the exit from strong supercooling is guaranteed because
the curvature of the temperature-dependent flaton potential at the origin eventually experiences a sign flip due to the negative curvature of the potential at zero temperature, i.e. the supercooling is guaranteed to be terminated. 
This termination of the supercooling is not limited to SUSY theories, and can arise in any theories in which a secondary mass scale appears in the effective potential of the scalar field, see e.g.~\cite{vonHarling:2017yew,Hambye:2018qjv,Schmitt:2024pby}.
This enforcement makes the transition fast and causes a subtlety about the end of the phase transition; does it end by 
nucleation and growth of the symmetry-breaking bubbles, or by the smooth change of the flaton expectation value which might be expected if the barrier is negligible? 

In Ref.\,\cite{Hiramatsu:2014uta}, it was argued that thermal fluctuations are already dominant compared to the bubble nucleation rate before bubbles can percolate.
At the ``would-be" percolation temperature obtained by calculating the bubble nucleation rate, the potential barrier is an order of magnitude smaller than the temperature, and therefore the scalar field is not trapped behind the potential barrier due to the large thermal fluctuations. 
The authors concluded that 
the scalar field can easily escape the meta-stable minimum and the phase transition must be finished by something akin to phase mixing, not by the formation of bubbles.
On the other hand, as shown in Ref.\,\cite{Langer:1967ax, Langer:1969bc, Langer:1974cpa, Kramers:1940zz, LINDE198137, 1983544, PhysRevLett.46.388, Berera:2019uyp, Gould:2021ccf}, the thermal escape from the meta-stable minimum is maximized at the critical bubble, i.e. the bounce solution.
Therefore, the smallness and (importantly) the shallowness of the potential barrier (which prevents fast run-away behaviour) should be already reflected in the value of the $O(3)$ bounce action $S_3/T$. 
This implies nothing but a very rapid nucleation of bubbles at the end of the phase transition. 
However, it is still subtle and uncertain whether a scalar field can be really trapped around the potential barrier whose size is smaller than the thermal fluctuations.

In this paper, we numerically simulate the Langevin equation with an appropriately flat representative potential\footnote{Due to subtleties when mapping a continuum theory onto a 3D lattice simulation, we do not directly model an exact flaton potential from a specific model but instead use a representative flat potential which captures the necessary features we wish to simulate.}, 
where the width of the potential barrier around the origin is smaller than the background temperature.
We explicitly show that the phase transition proceeds through the usual description of a first-order phase transition; bubbles nucleate, expand, and collide, rather than something akin to phase mixing.
Ultimately, the phase transition appears to be still described by the effective parameters as usual. 

This paper is organized as follows. Section~\Ref{Sec:mot} 
explains the setup we consider and two possibilities that describe the possible phase transition dynamics: bubble formation or phase mixing via growth of a tachyonic instability.
Section~\Ref{Sec:num_sim} describes the basics of our lattice simulation of a scalar field coupled to a thermal bath, and includes example qualitative snapshots of our simulations, for different assumptions of the potential shape given via benchmark models. 
Section~\ref{sec:SUSYpot} then discusses the basic properties of an example supercooled model and we discuss how this relates to our lattice simulations. 
As we conclude that the phase transition will proceed via bubble formation, we also provide some brief details about the possible gravitational wave spectrum.
Finally, in Section~\ref{sec:conclusion} we conclude.

\section{End of supercooled Universe}
\label{Sec:mot}

\subsection{
Thermal escape from supercooling
}

The scalar potential considered in this study is sufficiently flat in the sense that, for a given scalar field range $\Delta \phi$, the change in the scalar potential $\Delta V$ is much smaller than $(\Delta \phi)^4$. 
In particular, considering $v_\phi$ as the vacuum expectation value of $\phi$, we call a potential flat when the condition $V_0\equiv V(0) - V(v_\phi) \ll 10^{-2} \, v_\phi^4$ is satisfied where the factor of $10^{-2}$ comes from the characteristic structure of the one-loop thermal corrections. 
The Standard Model Higgs potential does not satisfy this condition as $\Delta V/v_h^4 \simeq 0.03$. 

\begin{figure}[t]
      \centering
      \includegraphics[width=0.8\textwidth]{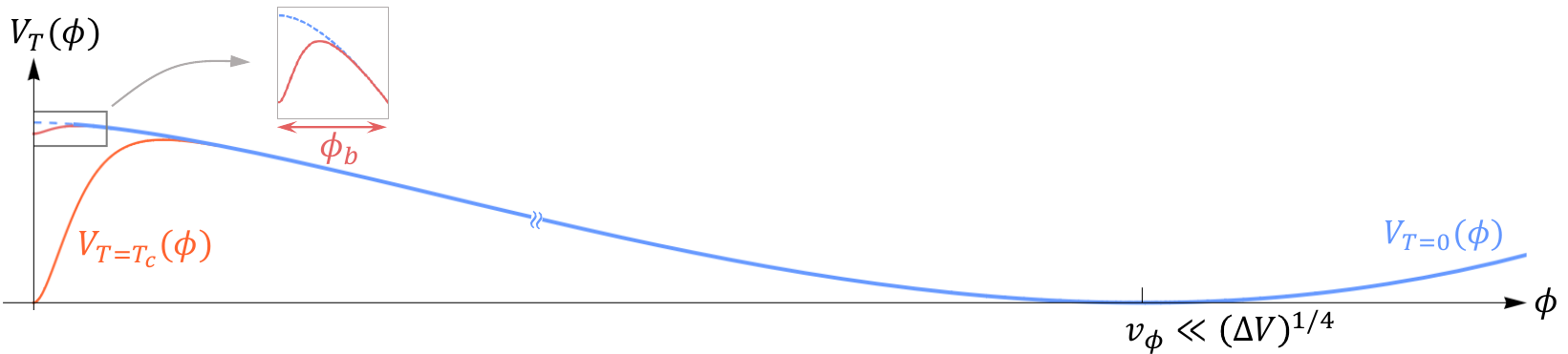}
      \caption{The temperature-dependent flaton potential at temperatures close to the phase transition. 
      } 
      \label{Fig:schematic_potential}
\end{figure}

A flat scalar potential in general leads to strong supercooling of the Universe, assuming that the initial temperature of the Universe is sufficiently large such that the symmetry that is broken spontaneously by the scalar field is restored.
As shown in Fig.~\ref{Fig:schematic_potential}, there is a large potential barrier between the symmetry-restored phase, $\langle \phi \rangle =0$, and the symmetry-breaking phase, $\langle \phi \rangle = v_\phi$, at the critical temperature, $T_c$, due to the flatness of the potential.
This makes the Universe remain trapped at the origin until the width and height of the potential barrier become comparable to the temperature scale. 
As a result, the Universe undergoes a long period of supercooling.

To make a transition from the meta-stable minimum to the true minimum, the scalar field must overcome the potential barrier, either through quantum or thermal fluctuations.
The quantum process occurs via tunnelling and the transition rate per unit volume is given in Ref.\,\cite{Coleman:1977py, Callan:1977pt}.
Although it may be important in some models, we restrict our situation to the case where the thermal escape rate dominates compared to the quantum tunnelling rate.

When the temperature fluctuation is large, 
the transition is mainly driven by thermal escape\,\cite{Langer:1967ax, Langer:1969bc, Langer:1974cpa, Kramers:1940zz, LINDE198137, 1983544, PhysRevLett.46.388, Berera:2019uyp, Gould:2021ccf}, which is explicitly demonstrated by the simulations in the next section.
The metastability of the potential around $\phi=0$ implies the system stays at the metastable minimum for a long time such that bubble nucleation can be treated as a locally rare event.

As shown in Ref.\,\cite{Berera:2019uyp, Gould:2021ccf}, the escape rate per unit volume can be derived from the Fokker-Planck equation, which is equivalent to the Langevin equation, and is expressed as
\bal
\label{Eq:Gamma_3}
\Gamma_3
=
A_{\rm dyn}
\Bigg( \frac{S_3}{2\pi T}  \Bigg)^{3/2}
\left[
\frac{{\rm det}[-\partial^2 + V''(0) ]}{\left|{\rm det'}[-\partial^2 + V''(\phi_3)]\right|}
\right]^{1/2}
\exp\left(- \frac{S_3}{T}\right),
\eal
where $S_3$ is the energy of the critical bubble configuration $\phi_3$ (explained below), $A_{\rm dyn}\sim \eta/2\pi $ is a dynamical prefactor, and
the prime in $\det$ denotes the removal of vanishing eigenvalues.
Here, $S_3$ must be understood as the minimally required energy of a scalar profile to make the transition, and the $\exp(-S_3/T)$ corresponds to the Boltzmann suppression factor in the configuration space of the scalar field.  

The escape from the local minimum can be initiated by a local thermal fluctuation
\bal
\phi(t_0,r)=0
\qquad
\xrightarrow[\text{fluctuation}]{\text{thermal}}
\qquad
\phi_3(r)+\epsilon(r),
\label{Eq:transition_3}
\eal
where $\epsilon(r)$ is a small and positive function of $r$ and $\phi_3(r)$ is the $O(3)$ bounce solution of
\bal
\label{Eq:eom_O3}
\frac{d^2 \phi_3}{dr^2} +\frac{2}{r} \frac{d\phi_3}{dr}  = \frac{\partial V_T(\phi_3)}{\partial\phi_3}
\eal
with the same boundary conditions as the $O(4)$ case.
We call such a transition as \emph{bubble nucleation} and the solution $\phi_3$ is referred to as the \emph{critical bubble}.
Notice that, even after the transition is made, $\phi(t,r)$ stays approximately static
if $\epsilon(r)=0$\footnote{One can find $d^2\phi/dt^2\simeq 0$ from the equation of motion with initial condition satisfying Eq.\,\eqref{Eq:eom_O3}}.
The bubble nucleation rate is defined as the rate of making the transition in Eq.~\eqref{Eq:transition_3} per unit volume.

\begin{figure}[t]
\centering
\includegraphics[width=0.5\textwidth]{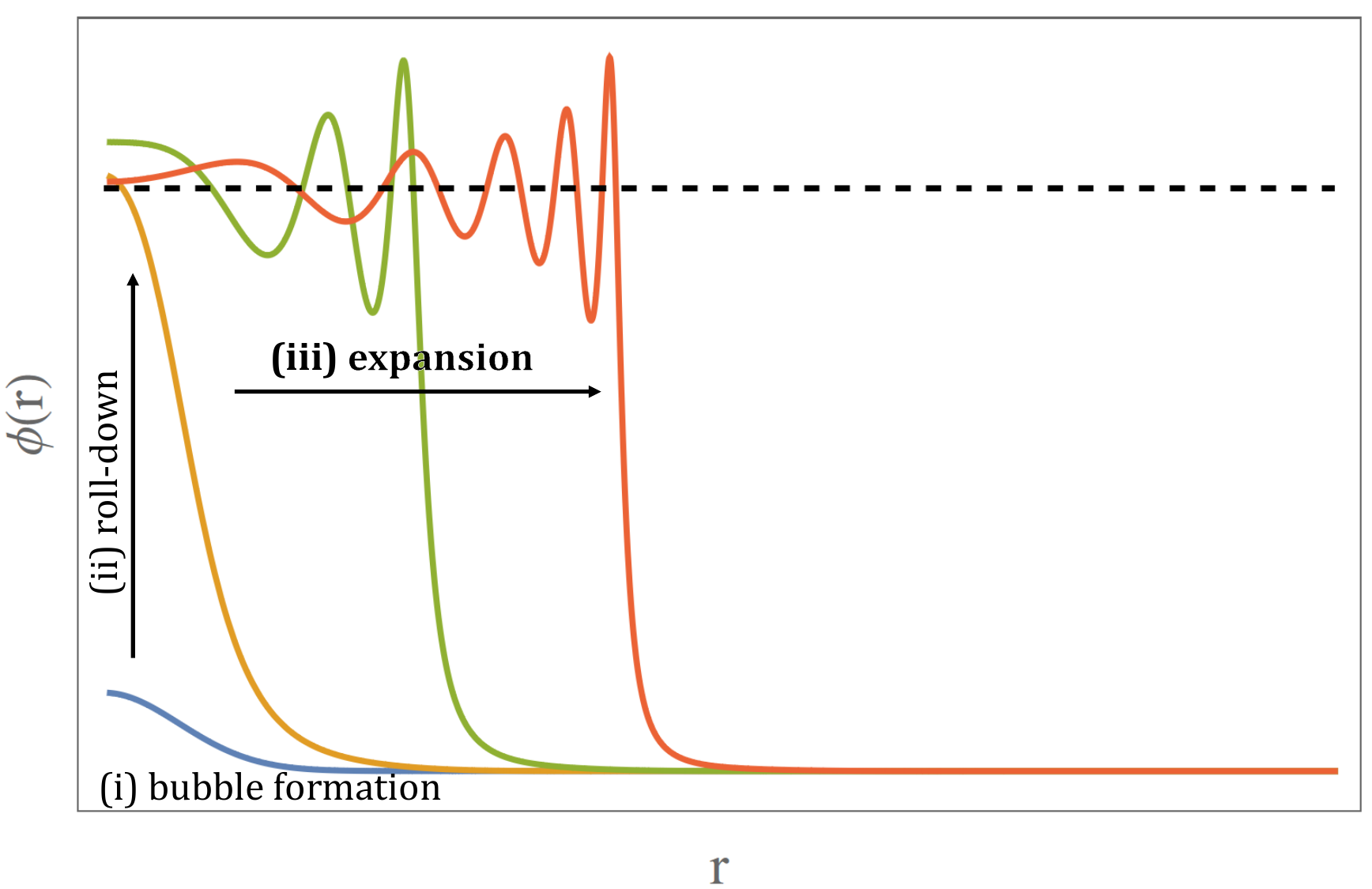}
\caption{Schematic plot of the field profile, in spherical coordinates, of a scalar field undergoing a first-order phase transition (either quantum or thermally induced). The three distinct phases are highlighted: (i) critical bubble formation, (ii) roll-down of the field to the true minimum, and (iii) localised expansion of this true-minimum bubble.}
\label{Fig:schematic_bubble}
\end{figure}

The process after a bubble nucleates undergoes three qualitatively distinctive steps as depicted in Fig.\,\ref{Fig:schematic_bubble}: (i) bubble nucleation, i.e. the transition in Eq.\,\eqref{Eq:transition_3}, (ii) the field around the centre of the bubble then rolling down to the true minimum, and (iii) localised bubble expansion centred around this nucleated bubble. This is schematically shown in Fig.~\ref{Fig:schematic_bubble}.

Let us estimate $S_3$ in terms of the potential parameters to understand the supercooling condition qualitatively (we obtain $S_3$ numerically later when a precise estimation is needed). Recall that $S_3$ represents the energy of the saddle point in the scalar field configuration space, which must be spatially spherically symmetric. We parameterize the characteristic radius of the profile of $\phi_3$ as $r_c$, the thickness of the wall as $\ell_w$, and the field difference as $\Delta \phi_c = \phi_c - 0$. The corresponding free energy difference is given by $\Delta V_c = V_T(0) - V_T(\phi_c) > 0$. A thin-wall bubble corresponds to the case where $\ell_w \ll r_c$, whereas when $\ell_w \sim r_c$, we refer to it as a thick-wall bubble. To obtain a more general expression, we treat $\ell_w$ and $r_c$ separately.  

The energy of the critical bubble consists of two contributions: the potential energy (a negative contribution) and the gradient energy (a positive contribution). The total energy of this profile is given by  
\begin{equation}  
E(r_c) = \int d^3 r \left[\frac{1}{2} \left(\frac{d\phi_3}{dr}\right)^2 + V_T(\phi_3(r)) - V_T(0) \right].  
\end{equation}  
The boundary conditions are $\phi_3(0) = \phi_c$ and $\phi_3(\infty) = 0$. Given that $\phi_3(r)$ smoothly transitions from $\phi_c$ to 0 within the range $\ell_w < r_c$, the energy can be approximately divided into two components:  
\begin{equation}  
E(r_c) \simeq  4\pi r_c^2 \ell_w \left[\frac{1}{2} \left(\frac{\Delta\phi_c}{\ell_w}\right)^2 + \Delta V_{\rm max} \right]
+
\int_0^{r_c} d^3 r \Big[V_T(\phi_c) - V_T(0) \Big] .  
\end{equation}  
Since the scalar field continuously transitions from the inside to the outside of the bubble, the potential also changes continuously while encountering a potential barrier. We denote the maximum potential difference at the bubble wall, relative to the potential at the core of the bubble, as $\Delta V_{\rm max}$, whose contribution is localized within the wall.  

With the tension of the bubble wall $\sigma_w$,  
\begin{equation}  
\sigma_w \sim \ell_w \left[ \frac{1}{2} \left(\frac{\Delta \phi_c}{\ell_w} \right)^2 + \Delta V_{\rm max} \right],  
\end{equation}  
the energy expression simplifies to  
\begin{equation}  
\label{Eq:E_rc}  
E(r_c) \simeq 
4\pi r_c^2 \sigma_w -\frac{4\pi r_c^3}{3} \Delta V_c .  
\end{equation}  
For small values of $r_c$, the gradient energy dominates, whereas for large $r_c$, the negative contribution from the potential energy becomes dominant. The radius $r_c$ is determined by maximizing $E(r_c)$ with respect to $r_c$, leading to  
\begin{equation}  
\label{Eq:r_c}  
r_c \sim \frac{\sigma_w}{\Delta V_c}.  
\end{equation}  
Substituting this into Eq.~\eqref{Eq:E_rc}, we obtain  
\begin{equation}  
\label{Eq:S3}  
S_3 \equiv E(r_c) \sim {\cal O}(10) \frac{\sigma_w^3}{(\Delta V_c)^2}.  
\end{equation}  

The equations of motion for the critical bubble profile describe the balance between the gradient energy and the potential energy stored in the bubble wall, which implies $ (\Delta \phi_c / \ell_w)^2 \sim \Delta V_{\rm max} $. From this, we find that the bubble wall tension is given by  
\begin{equation}  
\sigma_w \sim \Delta \phi_c \sqrt{\Delta V_{\rm max}}.  
\end{equation}  
Additionally, Eq.~\eqref{Eq:r_c} simplifies to  
\begin{equation}  
r_c \sim \frac{\Delta \phi_c \sqrt{\Delta V_{\rm max}}}{\Delta V_c},  
\end{equation}  
so that the ratio of $\ell_w$ to $r_c$ becomes  
\begin{equation}  
\frac{\ell_w}{r_c} \sim \frac{\Delta V_c}{\Delta V_{\rm max}}.  
\end{equation}  
Thus, we can identify that a thin-wall bubble is expected for $\Delta V_c \ll \Delta V_{\rm max}$, and the thick-wall case corresponds to $\Delta V_c \sim \Delta V_{\rm max}$.  

Applying these results to Eq.\,\eqref{Eq:S3}, we obtain 
\begin{equation} 
\frac{S_3}{T} \sim {\cal O}(10) \left( \frac{\Delta V_{\rm max}}{\Delta V_c} \right)^{3/2} \left( \frac{\Delta \phi_c}{T} \right) \left( \frac{(\Delta \phi_c)^4}{\Delta V_c} \right)^{1/2} . \label{Eq:S3overT_gen}
\end{equation}
Although $\Delta V_c$ and $\phi_c$ are dynamical variables, we can estimate them as $\Delta V_c \sim \Delta V_{\rm max}$ from the equation of motion unless the temperature is very close to $T_c$, and $\Delta \phi_c \sim \phi_b$ where the potential barrier ends at $\phi_b$ (see the inset box of Fig.\,\ref{Fig:schematic_potential}).
Here, our order estimation should not be understood as an equal sign, and a factor of a few is omitted; note that $\phi_c > \phi_b$.
With those approximations, we obtain
\bal
\frac{S_3}{T} \sim O(10) \left( \frac{\phi_b}{T} \right) \left( \frac{\phi_b^2}{\sqrt{\Delta V_{\rm max}}} \right).
\eal
This clearly implies that at least $\phi_b \sim T$ or $\phi_b \sim (\Delta V_{\rm max})^{1/4}$ is required such that bubble nucleation can be cosmologically effective, i.e. $S_3/T \lsim O(100)$.

For example, suppose the scalar potential at zero temperature takes the form  
\bal 
V(\phi) = V_0 + \kappa_4 \phi^4 \ln\left(\frac{\phi}{\mu_*}\right), 
\eal 
which has a minimum around $\langle\phi\rangle \sim \mu_*$. 
The flatness of the potential is achieved for \(\kappa_4 \ll 1\) which is plausible since $\kappa_4$ is generated radiatively. 
Taking $\Delta V_{\rm max} = \frac{\pi^2}{90} g_{*,\phi} T^4 \sim O(0.1)\,T^4$ with $g_{*,\phi}$ corresponding to the effective degrees of freedom that couples strongly to $\phi$. Solving $O(0.1)\, T^4 \sim \kappa_4 \phi_b^4 \ln(\phi_b/\mu_*)$, the ratio $\phi_b/T$ can be expressed as
\bal
\frac{\phi_b}{T} \sim \frac{1}{\, (O(10) \, \kappa_4 \,\log \mu_*/T )^{1/4}}.
\eal
Thus, $\phi_b/T \sim 1$ can be made only when $T \sim \mu_* \exp[-O(0.1)/\kappa_4] \ll \mu_*$, implying a strong supercooling (and possibly even eternal inflation if $\kappa_4$ is sufficiently small\,\cite{Levi:2022bzt,Marzo:2018nov,Lewicki:2021xku}).

Another relevant example (which is the main focus of this paper) is the scalar potential that can be expressed, at zero temperature, as
\bal 
\label{Eq:potential_msquared}
V(\phi) = V_0 - \frac{1}{2} m^2 \phi^2 + \cdots
\eal 
around $\phi=0$.
The potential has a global minimum at $v_\phi \gg m$ which can be realized by an additional contribution that ensures the flatness, e.g. supersymmetry. 
In the explicit model that we study in Sec.\,\ref{sec:SUSYpot}, this is expressed as $V_0\sim m^2 v_\phi^2 \ll v_\phi^4$.
With the inclusion of a thermal potential correction $\frac{c_2}{2} T^2 \phi^2 - \frac{c_3}{3} T \phi^3 - \frac{c_4}{4} \phi^4 \cdots $, 
the Universe can be trapped at the origin only when $ c_2 T^2 > m^2 $. 
Since there exists a temperature $T_2 = m/\sqrt{c_2}$ where the effective quadratic term, $-m^2+c_2 T^2$, flips its sign, we must separate two temperature regimes: $T \gg T_2$ and $T \sim T_2$.
When $T \gg T_2$, the effective potential is dominated by finite-temperature corrections, $\Delta V_{\rm max} \sim O(0.1)\, T^4$ is valid, and by solving $O(0.1)\,T^4 \sim m^2 \phi_b^2$, we obtain
\bal
\frac{\phi_b}{T} \sim \frac{T}{m} \sim \frac{1}{\sqrt{c_2}} \frac{T}{ T_2},
\eal
which is large. Thus, bubble nucleation can never be effective at $T\gg T_2$.
On the other hand, when $T\sim T_2$, the effective quadratic term becomes small.
Assuming $c_3$ dominates over $c_4$, the parameters $\phi_b$ and $\Delta V_{\rm max}$ in this case can be estimated as $(c_2 T^2 - m^2)/c_3 T$ and $\frac{5}{6} (c_2 T^2-m^2)^3/c_3^2 T^2$, respectively.
Thus, $S_3/T$ becomes
\bal
\frac{S_3}{T} \sim O(10) \frac{\sqrt{c_2}}{c_3} \sqrt{
\frac{T}{T_2}-1
},
\eal
and we conclude that bubble nucleation becomes effective only around $T\sim T_2$.
In fact, in many cases (like the model discussed in Section.\,\ref{sec:SUSYpot}), $c_3$ vanishes or is subdominant.
Repeating a similar estimation, one can find the same temperature dependence of 
\begin{equation}
\frac{S_3}{T} \sim O(10)\frac{\sqrt{c_2}}{c_4}\sqrt{\frac{T}{T_2}-1}
\label{eq:SUSY_S3T}
\end{equation}
when $c_3=0$, which will be used later in Section.\,\ref{sec:SUSYpot}.

This conclusion causes uncertainty in what really happens at the end of the phase transition.
First of all, $S_3/T$ changes rapidly around $T_2$, where bubble nucleation becomes cosmologically effective.
In this case, is the phase transition finished before the potential barrier disappears?
Secondly, $\phi_b/T$ becomes small even before the phase transition is finished, which implies that thermal fluctuations cause the field to not really see the local minimum at the origin (since $\Delta \phi \sim T$ from these fluctuations).
Thus, the potential that the scalar fluctuation effectively feels must be closer to that of a tachyonic potential, which would not involve the nucleation of bubbles.
We address these questions in this paper, but before going through the details, let us briefly discuss what happens when a scalar field undergoes a tachyonic instability.

\subsection{Thermal fluctuation in tachyonic instability}

In the symmetric phase, the expectation value of a scalar field coupled to a thermal bath
is given by $\langle \phi \rangle_T = 0$. 
However, these thermal noise contributions induce a nonzero variance for the scalar field, delocalising the field from the origin; this effect is sizeable when the potential curvature around the origin is smaller than the temperature of the bath. 
We begin by calculating the expected value of the two-point correlation function of $\phi$ for a scalar field with mass $m$ in thermal equilibrium as 
\bal 
\langle \phi({\bf x}) \phi({\bf y}) \rangle &= T \sum_{n=-\infty}^{\infty} \int \frac{d^3 {\bf k}}{(2\pi)^3} \frac{ e^{ i {\bf k} \cdot ({\bf x} - {\bf y})}}{\omega_n^2 + k^2 + m^2} \nonumber\\
&= \frac{m}{4\pi^2} \frac{K_1(mr)}{r} + \frac{m}{2\pi^2} \sum_{n=1}^{\infty} \frac{K_1 (m\sqrt{r^2 + n^2/T^2})}{\sqrt{r^2 + n^2/T^2}},
\eal 
where $r = |{\bf x} - {\bf y}|$ and $\omega_n = 2\pi n T$. Since $K_1(x) \sim 1/x$ for $x \to 0$, when $m \ll T$, we have
\bal 
\langle \phi({\bf x}) \phi({\bf y}) \rangle \big|_{m \ll T \ll 1/r} = \frac{\coth(\pi r T)}{4\pi r T} = \frac{1}{4\pi^2 r^2} + \frac{T^2}{12} + \cdots.
\eal 
The first term represents the zero-temperature value arising from the quantum nature of the scalar field. The variance of the scalar field due to thermal noise is then given by
\bal
\langle \phi^2({\bf x}) \rangle_T \simeq \frac{T^2}{12}.
\eal
The real-time dynamics of the scalar field, with thermal effects included, can be described by the classical Langevin equation for \( \phi \), as derived from the fluctuation-dissipation theorem:
\bal 
\partial_t^2 \phi({\bf x}, t) + \eta \partial_t \phi({\bf x}, t) - \nabla^2 \phi({\bf x}, t) + \partial_\phi V_T(\phi) = \xi({\bf x}, t),
\label{eq:sec2langevin}
\eal
where \( \xi({\bf x}, t) \) represents a stochastic noise term satisfying the following statistical properties:
\bal 
\langle \xi({\bf x}, t) \rangle_T = 0, \quad \langle \xi({\bf x}, t) \xi({\bf x}', t') \rangle_T = D \delta^{(3)}({\bf x} - {\bf x}') \delta(t - t'), 
\quad D = 2T\eta. 
\label{eq:sec2fluc-diss}
\eal
Here, \( \eta \) is a damping coefficient, associated with the dissipation of energy due to interactions with the thermal bath, typically of order the temperature \( T \), and \( V_T(\phi) \) is the thermal potential that accounts for temperature-dependent corrections. The Langevin equation describes the evolution of the field \( \phi \) under its own equations of motion as well as thermal noise and dissipation.

In general $\partial_\phi V_T(\phi)$ contains terms non-linear in $\phi$, making it difficult to evaluate analytically. However, if the potential is well approximated as
\bal 
V_T(\phi) \sim \frac{1}{2} m_T^2 \phi^2,
\eal
around the origin, and $|m_T|^2 \ll T^2$ is valid across the field range
$\Delta \phi \lesssim T$, 
we can solve the equation analytically. 
For the Fourier components of the field and the stochastic noise
\bal 
\phi({\bf x}, t) = \int \frac{d^3 {\bf k}}{\sqrt{(2\pi)^3}} \, \phi_{\bf k}(t) e^{i {\bf k} \cdot {\bf x}}, \quad \xi({\bf x}, t) = \int \frac{d^3 {\bf k}}{\sqrt{(2\pi)^3}} \, \xi_{\bf k}(t) e^{i {\bf k} \cdot {\bf x}},
\eal 
the Langevin equation becomes
\bal 
\ddot{\phi}_{\bf k}(t) + \eta \dot{\phi}_{\bf k}(t) + q^2 \phi_{\bf k}(t) = \xi_{\bf k}(t),
\eal where $q^2 = k^2 + m_T^2$, and
\bal 
\langle \xi_{\bf k}(t) \xi_{\bf k'}(t') \rangle_T = D \delta^{(3)}({\bf k} + {\bf k'}) \delta(t - t').
\eal 
For the initial conditions $\phi_{\bf k}(0) = \dot{\phi}_{\bf k}(0) = 0$, the analytic solution for $\phi_{\bf k}(t)$ yields the correlation function
$\langle \phi_{\bf k}(t) \phi_{\bf k'}(t) \rangle_T = D \delta^{(3)}({\bf k} + {\bf k'}) P_{\bf k}(t)$ 
with
\bal
P_{\bf k}(t) = \frac{1}{2 q^2 \eta} - \frac{e^{-\eta t} \left(\eta \cosh(\sqrt{\eta^2 - 4q^2} t) + \sqrt{\eta^2 - 4q^2} \sinh(\sqrt{\eta^2 - 4q^2} t) - 4q^2 / \eta \right)}{2 q^2 (\eta^2 - 4q^2)}.
\eal
This contributes to the scalar field variance in real space as

\[
\langle \phi^2({\bf x}) \rangle_T = 2 T \eta \int \frac{d^3 {\bf k}}{(2\pi)^3} P_{\bf k}(t)
\]
and when $m_T^2 \geq 0$, we have
\bal 
P_{\bf k}(t \to \infty) \simeq \frac{1}{2 q^2 \eta}.
\eal
The contribution up to the 
{\it cut-off} for the thermal fluctuation, $k \lesssim T$, gives $ \langle \phi^2({\bf x}) \rangle_T \sim T^2$.

If instead the potential resembled a Mexican hat potential around the origin, averaged over $\Delta \phi \sim T$, then $m_T^2 < 0$. In this case, $q^2 < 0$ for $k < |m_T|$, leading to a growing mode solution:
\bal 
 \langle \phi^2({\bf x}, t) \rangle_T \simeq \frac{T |m_T|}{8} \left(\frac{\eta}{2 \pi |m_T|^2 t}\right)^{3/2} \exp\left(\frac{ 2 |m_T|^2 t}{\eta}\right)
 \label{eq:sec2_tachyonic}
\eal for $|m_T| \ll \eta \ll |m_T|^2 t$. 
The result is an exponential increase in the variance with time, due to the tachyonic instability. 
Note however that, compared to the evaluation at zero temperature, the exponent is suppressed by the large thermal friction: $|m_T| / \eta\ll 1$. 
The growing field variance leads to a smooth phase transition, resembling phase mixing, rather than nucleation and growth of true vacuum pockets.

Thus, at the end of the phase transition, for the $m^2$-type flat potential in Eq.~\eqref{Eq:potential_msquared}, there are two possible transition channels which compete: the \emph{local} escape by bubble nucleation and the (approximately) \emph{global} runaway (which we refer to as phase mixing below) due to the tachyonic instability.
In a realistic scenario however, the flaton potential is not a simple Mexican hat potential, and it instead contains a local minimum at the origin while the phase transition completes. 
Therefore, it is subtle how to estimate the local and global escape rates analytically, and we turn to numerical simulations for clarification about the qualitative behaviour of such phase transitions.

\section{Numerical simulation}
\label{Sec:num_sim}
\subsection{Setup}
The aim of our numerical simulation is to explore the conditions that determine whether the phase transition is driven by bubble nucleation or by phase mixing.
For this, 
we consider a $\mathbb{Z}_2$-symmetric scalar potential, where the zero-temperature vacuum expectation value 
of the scalar field $\phi$
spontaneously breaks the $\mathbb{Z}_2$ symmetry ($\phi \to -\phi$). Then as the temperature decreases, the finite-temperature effective potential $V_T(\phi)$ induces a phase transition from $\phi = 0$ to $|\phi|= \phi_{\rm vev} >0$. Inspired by certain supercooled models,  a representative form of the effective potential we simulate is
\begin{equation}
\label{eq:4d-pot}
    V_{T}(\phi) = \frac{1}{2}m_T^2 \phi^2 - \frac{\lambda_T}{4!} \phi^4 + \frac{\epsilon}{6!} \phi^6.
\end{equation}
For example, this type of potential can arise in SUSY theories coupled to a thermal bath, where an exactly flat direction of the potential has a tachyonic instability due to SUSY breaking effects and is stabilised by higher-order terms. 
The parameters in Eq.~\eqref{eq:4d-pot}
should be understood as simulation parameters with thermal effects included, as the zero-temperature theory has a negative curvature at the origin.
Since the width of the potential barrier, in field space, is one of the key features at the onset of the phase transition, we fix the quadratic term by
\begin{equation}
\label{eq:pot_parameterised}
    m_T^2 = \frac{\lambda_T}{12} \phi_b^2,
\end{equation}
so that there is a potential barrier between from $\phi = 0$ up to $\phi = \phi_b$. 
By choosing $\phi_b = \alpha T$, the phase transition dynamics for a scalar field with a narrow potential barrier ($\alpha < 1$) can be simulated from an initial configuration at the meta-stable minimum.
The renormalisable parameters, $\phi_b$ and $\lambda$, effectively vary the bounce action, $S_3/T$, while $\epsilon$ adjusts the potential energy difference $\Delta V_T$ between the true and false vacua.
We expect that simulating different potentials, e.g., $\kappa_2 \phi^2 - \kappa_3 \phi^3 + \kappa_4 \phi^4$, will not change our qualitative conclusions for similar values of $\phi_b$, $S_3/T$, and $\Delta V_T$.

The extrema of Eq.~\eqref{eq:4d-pot} are given by
\begin{equation}
    \frac{\partial V_T}{\partial \phi} = 0 \implies \phi \in \{0,\,\phi_{\mathrm{top}},\,\phi_{\mathrm{vev}}\}
\end{equation}
where $\phi_{\mathrm{top}} = \phi_b/\sqrt{2}$ and $\phi_{\mathrm{vev}} \simeq 2\sqrt{5}\sqrt{\lambda_T/\epsilon}$, provided that $\epsilon$ is small: $\epsilon \phi_b^2 \ll \lambda_T$ (remember that $[\epsilon] = -2$). The height of the barrier
, $V_\mathrm{top} \simeq 10^{-2} \lambda \, \phi_b^4$, will also generically much smaller than $T^4$ unless the potential barrier width was taken very wide: $a\gg 1$.

Following the literature~\cite{Farakos:1994xh, Borrill:1994nk, Yamaguchi:1996dp, Borrill:1996uq, Cassol-Seewald:2007oak, Hiramatsu:2014uta, Gould:2024chm, Pirvu:2024ova, Pirvu:2024nbe}, the dynamics of the scalar field $\phi$ 
in the presence of thermal fluctuations
can be modeled through the 
the Langevin equation\footnote{In principle a multiplicative noise term $\xi_m(\mathbf{x}) \phi(\mathbf{x})$ should also be included but appears only to affect the time scale for the system to equilibrate. This, however, will complicate the implementation of the simulation~\cite{Cassol-Seewald:2007oak} so we neglect this contribution for simplicity.} defined in Eq.~(\ref{eq:sec2langevin}).
$\eta$ is understood as a damping coefficient and $\xi$ as a stochastic noise term assumed to be well modeled by uncorrelated, white noise as in Eq.~\eqref{eq:sec2fluc-diss}.
Such an assumption will remain valid as long as 
the lattice spacing of the simulation
is larger than the spatial and temporal correlation lengths of the noise, generated by the fermions and bosons within the thermal bath. This is  typically $\ell \sim (\pi T)^{-1}$ for fermions and exponentially damped for bosons~\cite{Yamaguchi:1996dp}.
The fluctuation-dissipation relation, $D=2T\eta$, 
ensures that the equilibrium values for $\phi$ do not depend on the damping parameter $\eta$, which only serves to control the time scale to equilibration. 

We simulate a 3-dimensional effective field theory containing the light bosonic field (the zero Matsubara mode), which we write as $\phi$
, with the remaining heavy fields integrated out~\cite{Farakos:1994xh}. The parameters within the dimensionally reduced theory, from herein 
denoted with the subscript $3$,
are related to the $3 + 1$ dimensional thermal theory through powers of temperature. In our case,
\begin{align}
    m_3^2 = m_T^2,\quad \lambda_3 = \lambda_T T \quad   \text{and} \quad  \epsilon_3 = \epsilon T^2
\end{align}
which carry appropriate mass dimensions. 
In our numerical simulations, we express all dimensionsful parameters relative to temperature, i.e. setting $T=1$, and it should be understood that all dimensionful quantities appear with appropriate powers of $T$ when applicable.

In order to solve~\eqref{eq:sec2langevin}, we discretise the dimensionally reduced 3-dimensional theory, $dx \rightarrow a$ and $dt \rightarrow \Delta t$ and numerically evolve each lattice point. For the spatial derivatives, we take
\begin{equation}
    \nabla^2 \phi(\mathbf{x}) = \frac{1}{12 a^2} \sum_i \left[ -\phi(\mathbf{x}-2i) + 16 \phi(\mathbf{x}-i) -30\phi(\mathbf{x}) + 16 \phi(\mathbf{x}+i) - \phi(\mathbf{x}+2i) \right]
\end{equation}
where we have assumed an improved form for $\nabla^2$ with $\mathcal{O}(a^4)$ accuracy.

    The non-smooth noise term within~\eqref{eq:sec2langevin} is discretised by simply replacing it with 
\begin{equation}
    \langle \xi(\mathbf{x},t) \xi(\mathbf{x'}, t') \rangle_T = D \delta(t-t')\delta^3(\mathbf{x} - \mathbf{x'}) \overset{\mathrm{lattice}}{\rightarrow} \frac{D}{(\Delta t) a^3}\delta_{x_i, x_j}\delta_{t_a,t_b}
\end{equation}
which can be simulated at each lattice point in space and time with a series of independent Gaussian-normal random variables
\begin{equation}
    \xi(\mathbf{x}_i, t_a) = \sqrt{\frac{D}{\Delta t a^3}} \mathcal{G}_{i,a}.
\end{equation}
Such choices ensure the correct variance for $\xi$ when mapping from the continuum theory to the lattice, however the order of convergence will now be at most $\mathcal{O}(a)$~\cite{telatovich2020strongconvergenceoperatorsplittingmethods}.

When mapping between a continuum and lattice field theory, the presence of thermal fluctuations leads to divergent behavior intimately related to the UV cut-off scale of the lattice simulation, $\Lambda_{\mathrm{UV}} \propto a^{-1}$. In order to properly match the equilibrium theory, lattice renormalisation counterterms should be included. These counter terms can be derived using a combination of dimensional analysis and lattice perturbation theory in order to derive the explicit values of $\delta X$ where the renormalised continuum parameters now appear on the lattice through $\delta X_{\mathrm{latt}} = X_{\mathrm{cont}} + \delta X$ where $X$ is coupling constants such as $m$, $\lambda$ and $\epsilon$~\cite{Gould:2024chm}.

The lattice discretised simulation of eq.~\eqref{eq:sec2langevin} is now performed with the replacements
\begin{align}
    V_T(\phi) \rightarrow V_3 (\phi) = \frac{1}{2}Z_\phi Z_m (m_3^2 + \delta m_3^2) \phi^2 &- \frac{1}{4!}Z_\phi^2(\lambda_3 - \delta \lambda_3) \phi^4 + \frac{1}{6!}Z_\phi^3(\epsilon_3 + \delta \epsilon_3) \phi^6\\
    \nabla^2 \phi &\rightarrow Z_\phi \nabla^2 \phi.
\end{align}
The lattice renomalisation terms have been calculated in some cases up to $\mathcal{O}(a^2)$~\cite{Moore:2001vf, Arnold:2001ir, Sun:2002cc} for $\phi^4$ theories which we modify with relevant contributions from $\epsilon_3$\footnote{
We however note that these additional contributions are not significant
for the case where $\epsilon_3$ is small.
The leading order contribution of $\epsilon_3$ only appears in $\delta \lambda_3$.
} which have leading order contributions given by
\begin{align}
    \delta m_3^2 &= -\frac{\Sigma \lambda_3}{8\pi a} + \mathcal{O}(\lambda_3^2 a^0)\\
    \delta \lambda_3 &= -\frac{\Sigma\epsilon_3 }{8\pi a} + \mathcal{O}(\epsilon_3^2 a^0) + \mathcal{O}(\lambda_3^2 a)\\
    \delta \epsilon_3 &= 0 + \mathcal{O}(\epsilon_3^2 a)\\
    Z_\phi &= 1 + \mathcal{O}(a^2)\\
    Z_m &= 1 + \mathcal{O}(a)
\end{align}
where $\Sigma$ is a renormalisation constant which has been calculated numerically and depends on the explicit choice for $\nabla^2 \phi$ used within the simulation. The exact form of these parameters which we use within our simulations, as well as the values of the various lattice renomalisation constants they depend on are neatly summarised in~\cite{Arnold:2001ir,Sun:2002cc,Gould:2024chm}.

\begin{figure}[t]
    \centering
    \begin{minipage}{0.4\textwidth}
        \includegraphics[width=\textwidth]{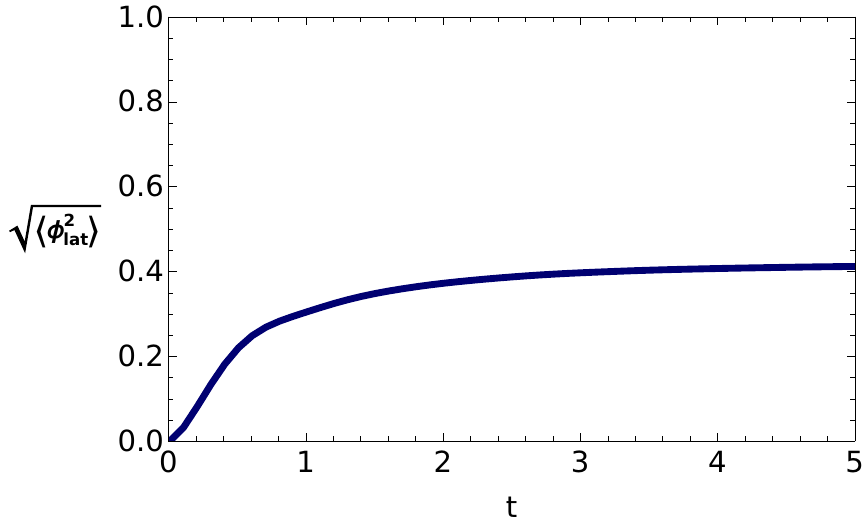}
    \end{minipage}
    \begin{minipage}{0.4\textwidth}
        \vspace{0cm}\includegraphics[width=\textwidth]{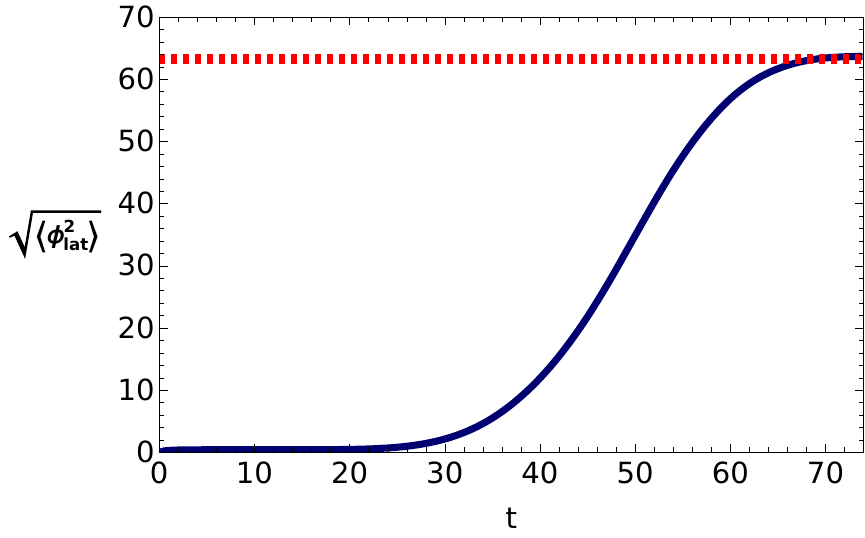}
    \end{minipage}
\caption{An example simulation of a phase transition from Benchmark A in Table~\ref{tab:c_H}. The left plot shows the initial thermalisation of the scalar field around the meta-stable origin very quickly starting from the initial condition $\phi(\mathbf{x},0) = \dot{\phi}(\mathbf{x},t) = 0$. After a hierarchically larger period of time, the system begins to transition to the true vacuum. The red-dashed line corresponds to the value of $\phi_{\rm vev}$ in the continuum theory.}
\label{fig:therm_time}
\end{figure}

As an initial condition for our simulation, we take
\begin{equation}
    \phi(\mathbf{x},0) = \dot{\phi}(\mathbf{x},t) = 0
\end{equation}
which, while unphysical, does not have any significant effect on the final results because the system rapidly reaches thermodynamic equilibrium around the meta-stable minimum. 
The time scale for this thermalisation is much shorter than the time scale of escaping from the metastable minimum.
This can be seen explicitly in Fig.~\ref{fig:therm_time}, where we depict the evolution of $\sqrt{\langle \phi^2 \rangle}$ with a representative choice of parameters (benchmark A in Table.\,\ref{tab:c_H}).
The left panel shows the time scale needed for the initial thermalisation (which results in a  $O(T)$ change in $\sqrt{\langle \phi^2 \rangle}$) while the right panel shows the time scale required for the whole phase transition of the system (the potential is minimized at the value marked by the red dotted line).

The time evolution of the simulation, which was performed in our Python code, utilises the fourth-order symplectic Forest-Ruth algorithm~\cite{FOREST1990105}. We have tested our simulations against other algorithm choices, such as second- or fourth-order Runge-Kutta as well as the second-order symplectic leapfrog method, and found no significant variation in the predictions for the phase transition.

The approach described above is based on several assumptions\,\cite{Kajantie:1995dw}. 
It assumes that the dynamics are dominated by wavelengths of $\phi$ which are much longer than that of the particles fluctuating in the thermal bath with which it couples to, and that the fluctuations of the thermal bath are in thermal equilibrium. 
Under these assumptions, all corresponding short-wavelength degrees of freedom ($|\mathbf{k}| \gtrsim \pi T$) are integrated out, leaving only the Matsubara zero mode of the scalar field, whose equation of motion is given by the Langevin equation \eqref{eq:sec2langevin} with a thermal effective potential $V_T$.

As supercooling implies a significant lowering of the temperature before bubble percolation, it may be natural to question whether the percolation temperature could be so low that the mass scales appearing are comparable to, or even lower than $T$, in tension with the conditions required above\footnote{This point has also been discussed in~\cite{Kierkla:2023von,Kierkla:2025qyz}}.
Focusing first on general properties of flat potentials which exhibit supercooling and have an epoch of curvature flipping: (i) the flatness of the potential enforces \(\sqrt{|V''(\phi)|} < T\) even for large field values\footnote{Take for example a scalar potential relevant for supercooling like \(-m^2\phi^2 + \phi^6/\Lambda^2\), even if $\phi_{\rm vev}$ is much larger than \(m\), the size of \(\sqrt{|V''(\phi)|}\) between the origin and $\phi_{\rm vev}$ remains of the order of \(m\) and is nearly independent of the value of \(\phi\).} and, (ii) the phase transition occurs around the temperature scale of curvature flipping where the barrier is rapidly disappearing, so at the origin \(0 < V_T''(0) = (-m^2 + cT^2) \ll T^2\). This large hierarchy justifies such an effective description.

Instead, in the case of the toy potential simulated on the lattice~(\ref{eq:4d-pot}), $V(\phi)$ scales as \(-\lambda_T \phi^4\) for \(\phi\) sufficiently far away from the end of the barrier, and thus \(\sqrt{|V_T''(\phi)|}\) will eventually become larger than the heavy thermal modes at some cut-off scale \(\phi_{\rm cutoff}\). Since the dynamics of critical bubble formation is determined through the properties of the potential barrier (and this potential is sufficiently flat in this range), the simulation of such a toy potential will be relevant in determining the nature of the phase transition (i.e. whether it is driven by bubble formation or not) provided that that critical bubble formation occurs sufficiently below $\phi_{\rm cutoff}$.
We show below that this is indeed the case for all of our benchmark points and therefore our simulations are a consistent test of bubble formation for supercooled potentials with small barriers. We also numerically verify that our lattice simulation results are heavily saturated by the long-wavelength modes of $\phi$ with a negligible population of high-wavelength modes.

Of course, this suggests that, at sufficiently large field values of $\phi>\phi_c$, a modification is required to account for the short wavelength modes becoming dynamically relevant, perhaps by extrapolating between the 3D and 4D potentials.
As in this regime a number of important effects have already been ignored, such as the inherent $\phi$ dependence of $\eta$ (fields coupled to $\phi$ eventually become heavy during the PT -- and therefore Boltzmann suppressed in the plasma), or accounting for the inhomogeneities of the fluid around the bubble wall\,\cite{Kamionkowski:1993fg, Espinosa:2010hh}. Such effects are relevant, e.g. an extraction of the bubble wall velocity and therefore bubble radii at collision, and accounting for such effects is beyond the current scope, which is specifically concerned about dynamics around the region of bubble formation.

\subsection{Benchmarks and simulation parameters}

A semi-analytic expression for the thermally-induced bounce action has been derived~\cite{Linde:1981zj} (while we have also checked it numerically) for Eq.~\eqref{eq:4d-pot} in the tachyonic regime, $\epsilon \rightarrow 0$:
\begin{equation}
\label{Eq:S3overT_simulation}
    \frac{S_3}{T} \simeq \frac{114\sqrt{m^2}}{\lambda T}\quad
    \xrightarrow{{\rm Eq.}~\eqref{eq:pot_parameterised}} 
     \quad \frac{S_3}{T} \simeq 19\sqrt{3} \frac{\phi_b}{\sqrt{\lambda} T}
\end{equation}
which will remain valid provided that  
the global minimum is far from the location of the potential barrier. 
This condition is anyway naturally satisfied in supercooled transitions. 
On the other hand, for small barriers $\phi_b < T$, there is a significant contribution to the bubble nucleation rate stemming from the complicated numerical prefactor appearing in the nucleation rate: $\Gamma_n \simeq A \exp (-S_3/T)$, which we evaluate numerically using BubbleDet~\cite{Ekstedt:2023sqc}.

\begin{table}[t]
    \centering
    \renewcommand{\arraystretch}{1.15}
\begin{tabular}{|c|c||c|c||c|c||c |c |}
    \hline
    & \{$\lambda$, $\phi_b$, $\epsilon$\} & $\phi_{\rm vev}/T$ & $\left|\Delta V_T/T^4\right|$ & $S_3/T$ & $\log \Gamma_n/T^4$ & $\frac{\sqrt{|V''(\phi_c)|}}{(2\pi T)}$ & $\frac{\phi_{\rm cutoff}}{\phi_{\rm c}}$\\
    \hline
    A & \{$2$, $0.3 T$, $T^2/100$\} & $63.26$ & $4.44 \times 10^5$ & $6.98$ & $-15.6$ & $0.145$ & $6.889$\\
    B & \{$2$, $0.3 T$, $10\, T^2$\} & $1.98$ & $0.4$ & $7.43$ & $-16.62$ & $0.114$ & $3.770$\\    
    C & \{$10$, $0.5 T$, $T^2/100$\} & $141.51$ & $5.55 \times 10^{7}$ & $5.17$ & $-9.01$ & $0.541$ & $1.879$\\    
    \hline
    \hline
    M & \{$-0.011$, $10.95\, T$, $0$\} & $7.74$ & $1.66$ & $-$ & $-$ & $-$ & $-$\\
    \hline
\end{tabular}
    \caption{Benchmark simulation parameters in terms of \{$\lambda$, $\phi_b$, $\epsilon$\} assuming Eqs.~\eqref{eq:4d-pot} and~\eqref{eq:pot_parameterised}. Although benchmarks B and C do not correspond to a weakly-coupled perturbative theory, we consider such simulations on the lattice in order to observe the behaviour of the nature of the phase transition as $S_3/T$ or $\Delta V_T$ changes. Benchmark M instead corresponds to a Mexican hat potential without a potential barrier and a relatively large negative curvature at the minimum, $m^2 = -T^2/9$; which we use to compare the simulation results to a case without bubble formation. 
    The two rightmost columns depict quantities to test the validity of our simulation, and all benchmark sets satisfy $\sqrt{|V''(\phi_c)|}/(2\pi T) <1$ and $\phi_{\rm cutoff}/\phi_c>1$, which justifies our framework. 
    }
\label{tab:c_H}
\end{table}

\begin{figure}[t]
    \centering
    \includegraphics[width=0.5\textwidth]{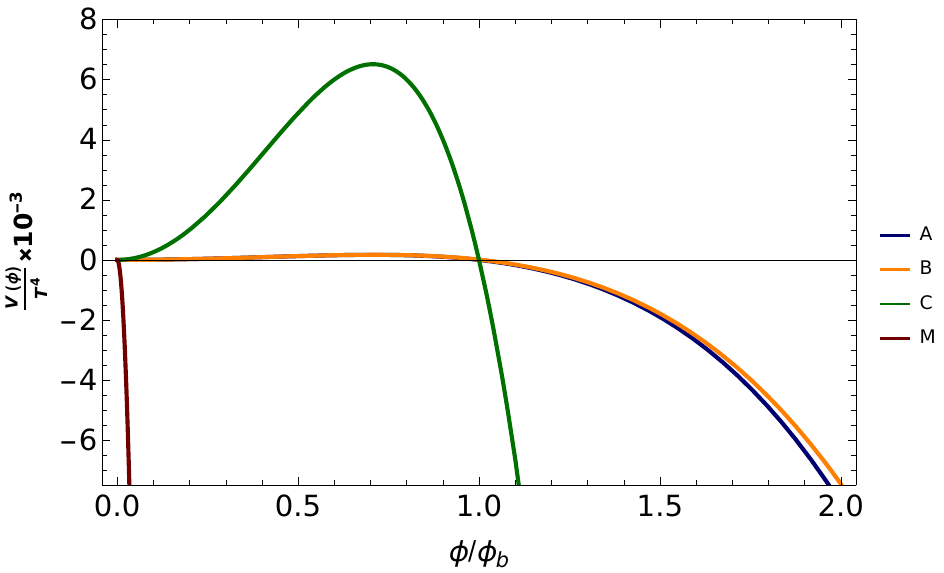}  
\caption{Comparative plot of the shape of the potential around the origin, normalised to their respective barrier widths $\phi_b$, for the benchmark points in Table~\ref{tab:c_H}. The shape of benchmark points A and B are unsurprisingly similar, whereas benchmark C steeply drops after the barrier disappears, leading to a larger nucleation rate. Benchmark point M contains no barrier and very steeply drops away from the origin compared to the other three cases.}
\label{fig:bench_barrier}
\end{figure}

We define a number of benchmark points, (A-C), in Table~\ref{tab:c_H} to demonstrate the results of our simulations. 
In these benchmarks, we fix $\phi_b$ to be smaller than the temperature scale associated with the thermal bath. 
This implies that the distribution of $\phi$, after it thermalises around the meta-stable minimum ($\langle \phi^2 \rangle \sim T^2$), extends beyond the barrier itself, and we observe the implications for the field as it eventually evolves down to the global minimum.
Details of other parameters are simply chosen
to represent the dependence in the qualitative behaviour of the phase transition as $S_3/T$, $\Gamma_n$, and $\Delta V_T$ vary. 
We also define a benchmark (M) to compare against a simulation in which exactly no barrier exists for a standard Mexican hat potential with a relatively large negative curvature at the origin.
Their respective potential shapes close to the origin are displayed in Fig.~\ref{fig:bench_barrier}. 
The flatness of the potential for benchmark points A and B around the origin is ultimately responsible for their much slower predicted bubble nucleation rate compared to benchmark point C, even though C has a much taller barrier (but nonetheless small compared to $T^4$).
This is consistent with Eq.\,\eqref{Eq:S3overT_simulation}: $S_3/T \propto 1/\sqrt{\lambda}$. 

The two rightmost columns in Table\,\ref{tab:c_H} depict quantities to test the validity 
of the dimensionally reduced simulation; first at the central field value of the critical bubble, $\phi_c$ (note that $\phi_c \simeq 3\phi_b$ for our toy potential), compared to the (thermal) mass of the first non-zero Matsubara mode, and second, the value of $\phi_{\rm cutoff}$ at which the first non-zero Matsubara mode and the zero mode have comparable masses, i.e. $\sqrt{|V''(\phi_{\rm cutoff})|} = 2\pi T$.
As we discussed in the earlier section, all benchmark parameters satisfy $\sqrt{|V''(\phi_c)|}/(2\pi T) <1$ and $\phi_{\rm cutoff}/\phi_c>1$, justifying our framework.

Benchmarks B and C may not seem reasonable from a model-building point of view due to the large values of $\epsilon$ or $\lambda$.
However, we stress that our simulation of these benchmarks remains instructive for cases where $\Delta V/T^4$ is small (B) or $\Gamma_n/T^4$ is large (C) respectively.
The
value of $\lambda=2$ for benchmark A, corresponding to a \emph{realistic} supercooled transition with a small potential barrier, was chosen purely pragmatically such that the simulation time for the phase transition,
\begin{equation}
    t_{\rm first-bubble} \sim \frac{1}{\Gamma_n L^3}
\end{equation}
is short enough to feasibly perform multiple simulations. 
For smaller values of $\lambda$, the simulation time is expected to significantly increase as $S_3/T \propto \lambda^{-1/2}$, which we also observe numerically. 
However, we remark that a full quantitative comparison of the simulation results to the analytic predictions, i.e. extracting $\Gamma_n^{\rm lattice}$ (as was done in e.g.~\cite{GRIGORIEV198867,PhysRevB.42.6614,PhysRevLett.68.1645,PhysRevD.47.R2168,Alford:1993ph,Borsanyi:2000ua,Batini:2023zpi,Gould:2021dzl,Gould:2024chm}), is beyond the scope of the current work and will require significantly more validation, but is planned for the future. 
In this work, we do not discuss the evaluation of $\Gamma_n^{\rm lattice}$, but we only focus on whether the phase transition is driven by the bubble nucleation or phase mixing.

As made clearer in Sec.~\ref{sec:SUSYpot}, 
a realistic model would have a much larger hierarchy between $\phi_{\rm vev}$ and $T$ than what is assumed in Table~\ref{tab:c_H}, corresponding to a much smaller value of $\epsilon/\lambda$.
However, as the thermally induced bounce action is independent of $\epsilon$ for sufficiently small values, the phase transition dynamics (in particular the time-scales associated with bubble nucleation) are independent of $\epsilon$. 

In our simulation, the time step $\Delta t$ is taken to be smaller than the characteristic time scale for oscillations around the global minimum 
\begin{equation}
    \tau \sim \frac{1}{\sqrt{V''_T(\phi_{\rm vev})}} \sim \frac{\sqrt{\epsilon}}{\lambda}.
\end{equation}
$\tau$ indeed corresponds to the time scale when the field is rolling down after a bubble has nucleated, and therefore this enforces the rough constraint $\Delta t < \tau$ to avoid pathological behaviour as the field oscillates around the true minimum.
The decoupling of the bubble nucleation dynamics from the location of the global minimum allows us to choose 
$\epsilon$ larger than a realistic value, and therefore 
a larger time step. This makes the simulation easier, 
without affecting any quantitative properties of the phase transition itself. We have verified that smaller values of $\epsilon$ do not have an impactful effect on the simulation results beyond the larger computation requirements.

\begin{figure}
    \centering
    \includegraphics[width=0.5\linewidth]{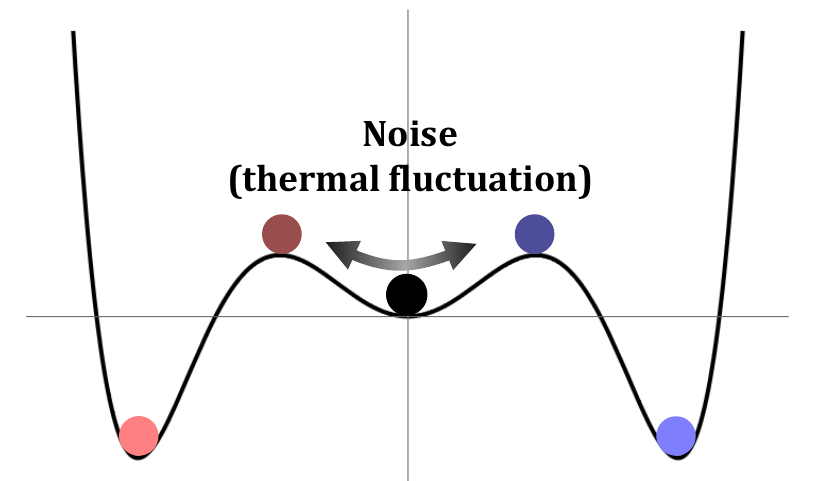}
    \caption{Schematic (not to scale) plot which shows the colour scheme for visualizing our simulation snapshots. The actual potential we simulate has a much smaller potential barrier compared to the global minimum, as shown in Fig.\,\ref{fig:bench_barrier}.}
    \label{Fig:color_scheme}
\end{figure}

We have performed the simulation several times for different choices of lattice parameters and observed consistent results.
Thus, in what follows, we depict our results only for a given choice of lattice parameters: \{$\Delta t$, $a$, $L$\}.

\subsection{Simulation results}

As indicative of the qualitative nature of the phase transition in each case, we present simple two-dimensional snapshots of the field value along a fixed $z$-slice at various specified points in time.
At each time step, we associate each lattice point with a colour which smoothly changes as the value of $\phi$ changes (see Fig.\,\ref{Fig:color_scheme}): 
(i) when $\phi$ is close to the origin, the lattice point is coloured black, 
(ii) as $\phi$ values move to be around the top of the potential barrier (including beyond the barrier) the lattice point colour transitions to a dark red or blue depending on which side of the origin, and 
(iii) as the field moves closer to the degenerate global minima the lattice points become a progressively lighter red or blue depending on which of the two minima it approaches  (recall that each minima of our potential breaks the $\mathbb{Z}_2$). 
However, whether the field remains localised around the origin until a bubble nucleates or whether it simply begins to roll down once it moves beyond the barrier is clearly distinguishable in this colour scheme.

\begin{figure}[tp]
    \centering
    \begin{minipage}{0.32\textwidth}
        \includegraphics[width=\textwidth]{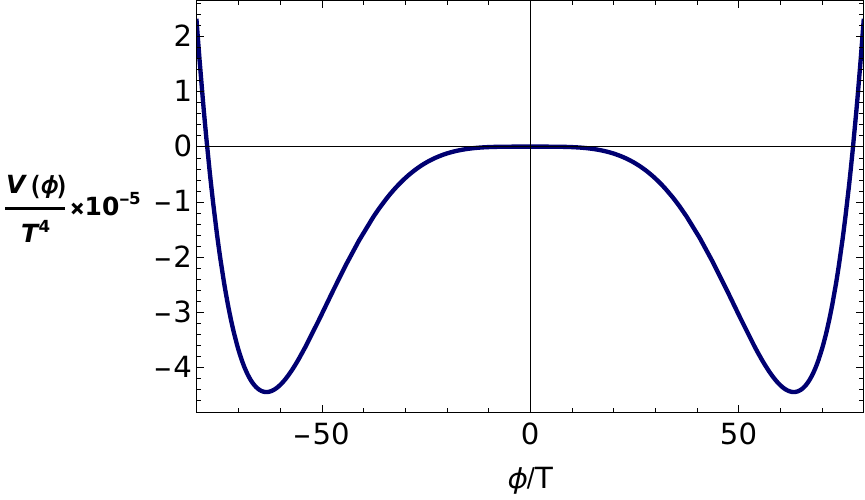}
    \end{minipage}
    \begin{minipage}{0.32\textwidth}        \vspace{-0.15cm}\includegraphics[width=\textwidth]{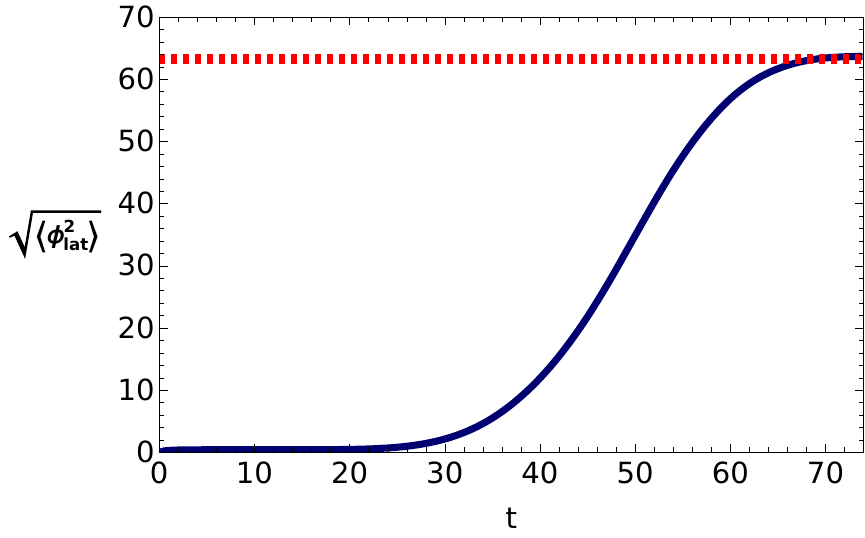}
    \end{minipage}
    
    \begin{minipage}{0.31\textwidth} 
        \includegraphics[width=\textwidth]{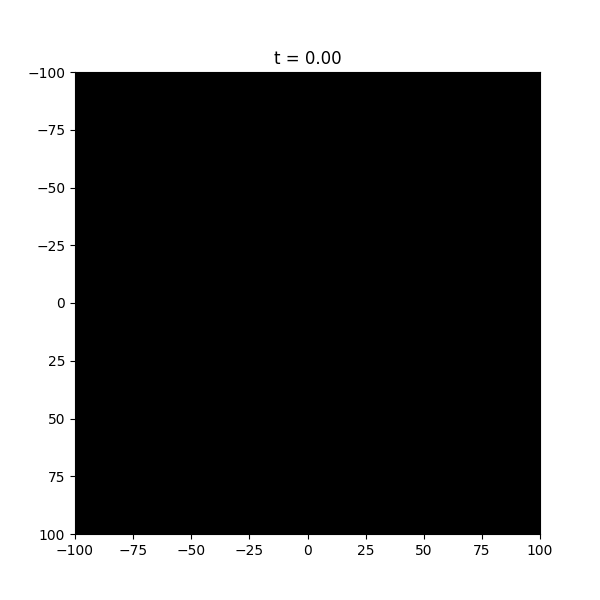}
    \end{minipage}
    \begin{minipage}{0.31\textwidth}
        \includegraphics[width=\textwidth]{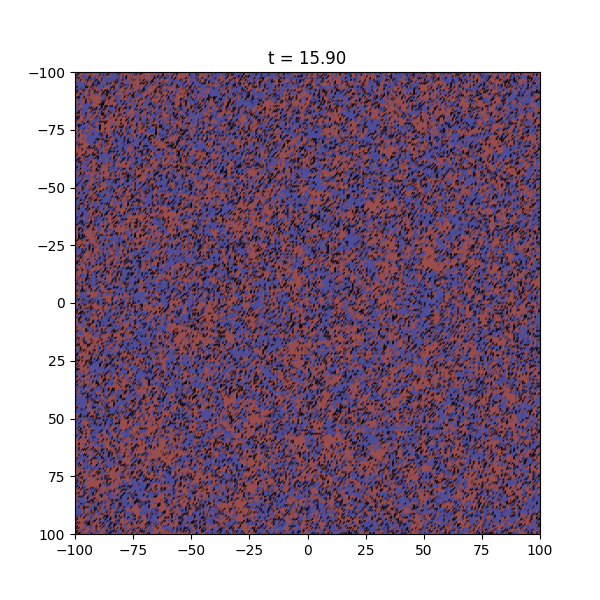}
    \end{minipage}
    \begin{minipage}{0.31\textwidth}
        \includegraphics[width=\textwidth]{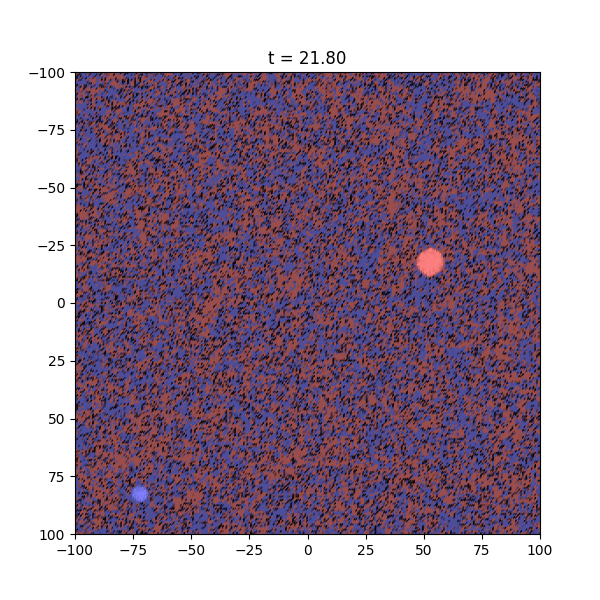}
    \end{minipage}
    
    \begin{minipage}{0.31\textwidth} 
        \centering
        \includegraphics[width=\textwidth]{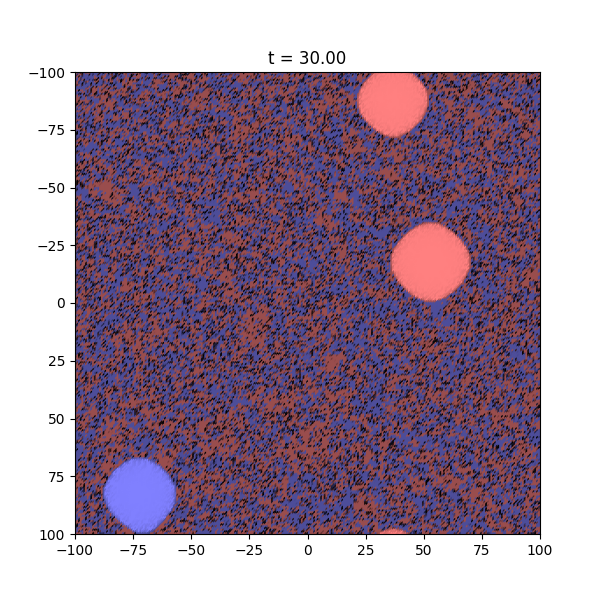}
    \end{minipage}
    \begin{minipage}{0.31\textwidth}
        \centering
        \includegraphics[width=\textwidth]{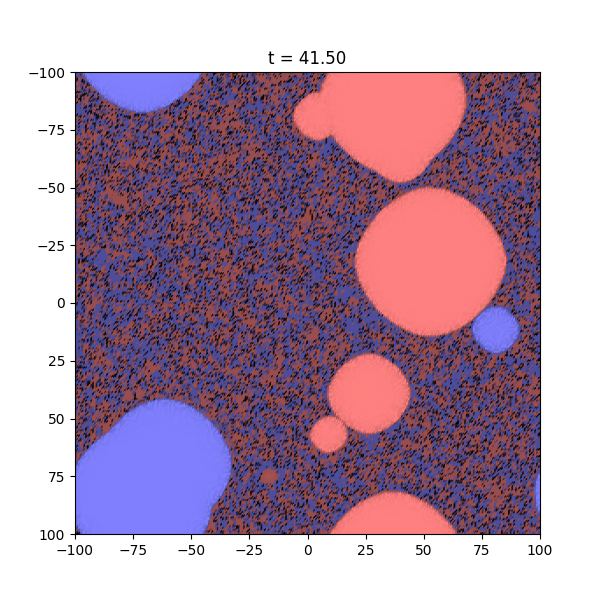}
    \end{minipage}
    \begin{minipage}{0.31\textwidth}
        \centering
        \includegraphics[width=\textwidth]{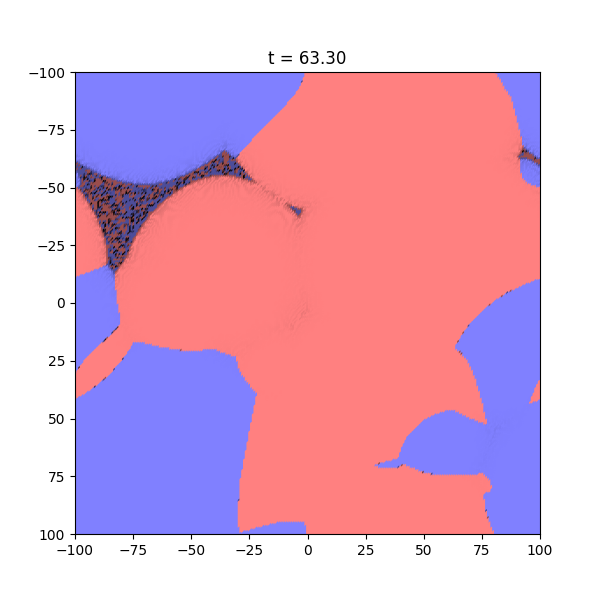}
    \end{minipage}    
    \caption{Benchmark A: \textbf{(Top)} Plot of the potential shape (left) and evolution of the average field value on the lattice, $\sqrt{\langle \phi^2 \rangle }$ (right) along with the continuum prediction for $\phi_{\rm vev}$.
    \textbf{(Below)} Snapshots of the lattice simulation in one $z$-slice. black points correspond to lattice points with field values close to the meta-stable minimum, dark red or blue points correspond to points around (and beyond) the potential barrier, and light red and blue field values are lattice points with field values close to the (degenerate) global minima. The simulation results indicate a phase transition proceeding via bubble formation and nucleation.}
    \label{fig:benchA}
\end{figure}

Fig.~\ref{fig:benchA} depicts the evolution of the lattice simulation, along with the potential shape, at different snapshots in time for benchmark point A. The scalar field, with homogeneous initial condition $\phi(\mathbf{x},0) =0$, quickly thermalises around the origin with a fixed variance, as shown on the left-side of Fig.~\ref{fig:therm_time}, $\langle \phi^2 \rangle \sim T^2 > \phi_b^2$. 
This localisation is a result of the gradient energy in Eq.~\eqref{eq:sec2langevin} which prevents the field from simply random walking with time away from the origin. Although a significant fraction of the field distribution lies beyond the barrier, for a large period of time the field is unable to roll down to the minimum. Eventually, bubbles begin to nucleate and grow with time, so the result is a first-order phase transition via bubble nucleation. The different regions of red and blue are delineated by domain walls which is a result of the $\mathbb{Z}_2$ symmetric potential we have simulated.

\begin{figure}[tp]
    \centering
        \begin{minipage}{0.32\textwidth}
        \includegraphics[width=\textwidth]{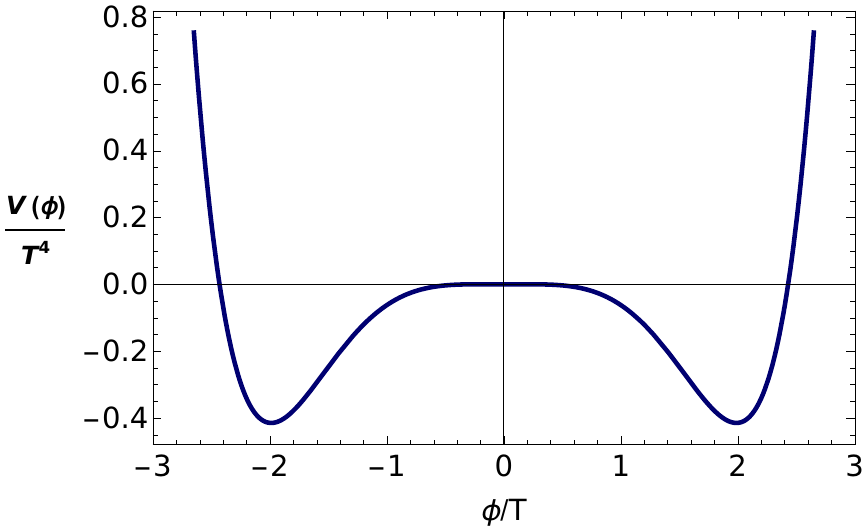}
    \end{minipage}
    \begin{minipage}{0.32\textwidth}        \vspace{-0.15cm}\includegraphics[width=\textwidth]{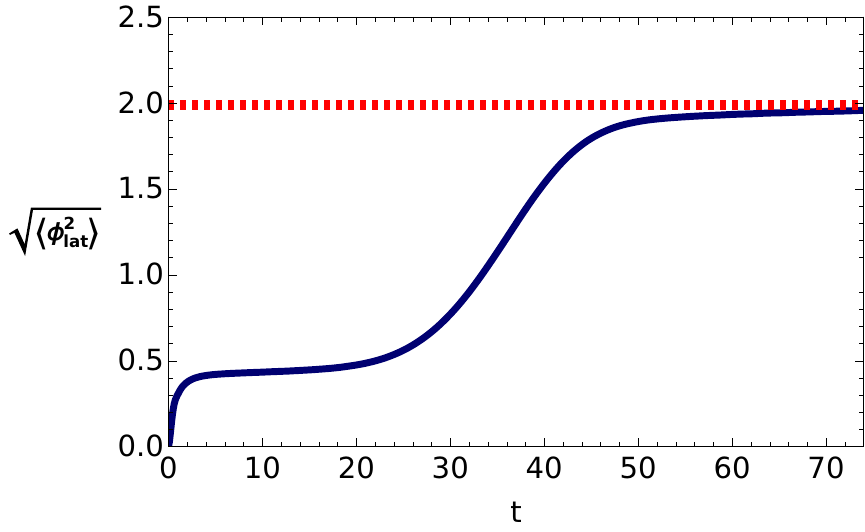}
    \end{minipage}    
    
    \begin{minipage}{0.31\textwidth} 
        \includegraphics[width=\textwidth]{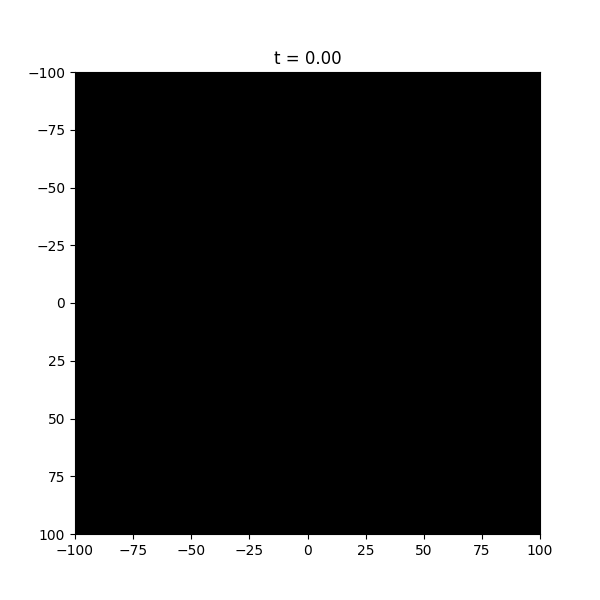}
    \end{minipage}
    \begin{minipage}{0.31\textwidth}
        \includegraphics[width=\textwidth]{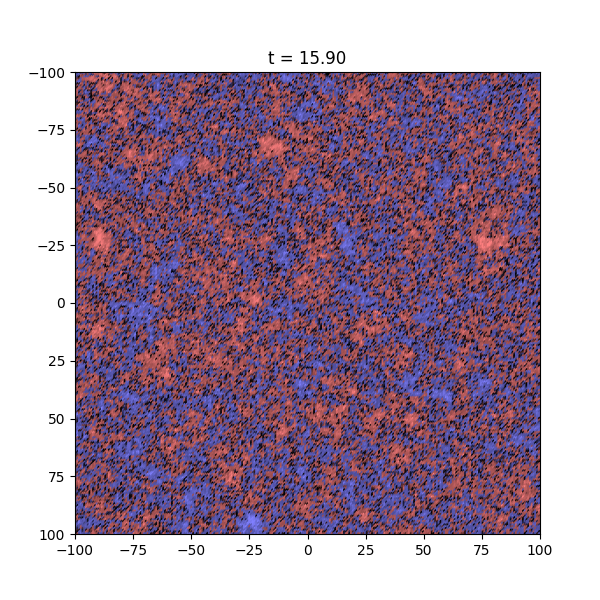}
    \end{minipage}
    \begin{minipage}{0.31\textwidth}
        \includegraphics[width=\textwidth]{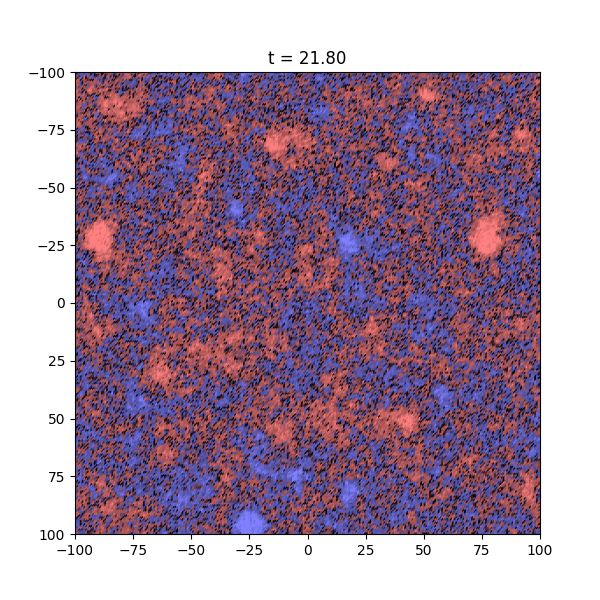}
    \end{minipage}
    
    \begin{minipage}{0.31\textwidth} 
        \centering
        \includegraphics[width=\textwidth]{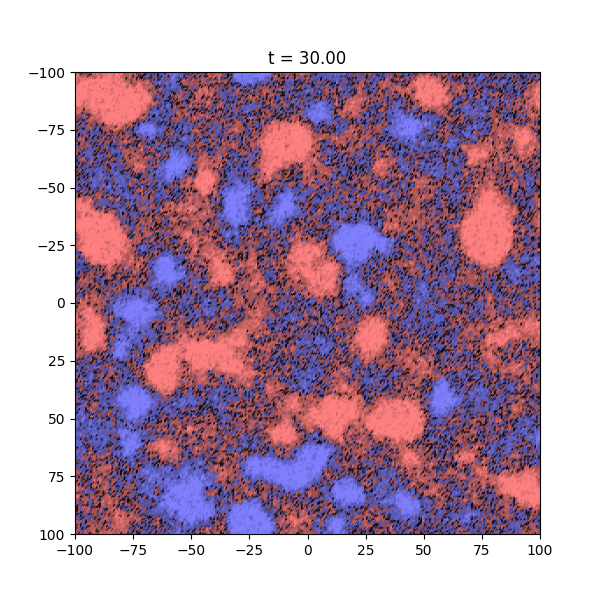}
    \end{minipage}
    \begin{minipage}{0.31\textwidth}
        \centering
        \includegraphics[width=\textwidth]{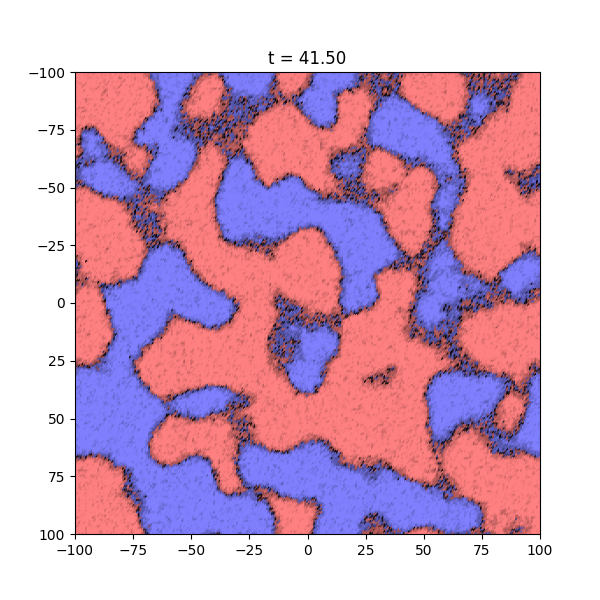}
    \end{minipage}
    \begin{minipage}{0.31\textwidth}
        \centering
        \includegraphics[width=\textwidth]{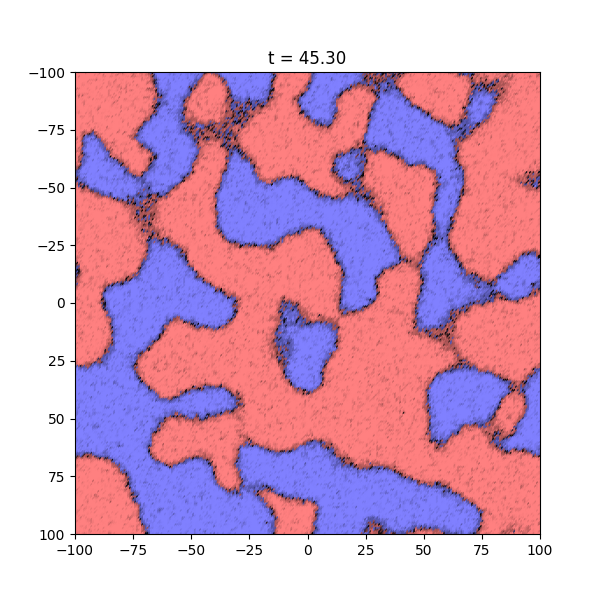}
    \end{minipage}   
    \caption{Benchmark B: \textbf{(Top)} Plot of the potential shape (left) and evolution of the average field value on the lattice, $\sqrt{\langle \phi^2 \rangle }$ (right).
    \textbf{(Below)} Result of the lattice simulation in one $z$-slice at different time steps. The simulation results indicate a phase transition in a first-order-like manner however the `bubbles' seem to instead be significantly affected by thermal fluctuations due to the small value of $\Delta V$ in the simulation. Note that the timescales of this phase transition appear to coincide with that of Fig~\ref{fig:benchA}.}
    \label{fig:benchB}
\end{figure}

The evolution of Benchmark B is displayed in Fig.~\ref{fig:benchB}. As indicated in Table~\ref{tab:c_H}, the predicted bubble nucleation properties for cases A and B are almost the same, so the time scale between the two snapshots appears to be roughly of the same order. However, unlike case A, there is no significant hierarchy between the location of the global minimum or the potential difference when compared to the temperature scale. As a result of $\Delta V_T/ T^4 < 1$, the would-be symmetric bubbles appear to receive significant deviations in their geometry from the thermal fluctuations induced by the bath. Nonetheless, the phase transition appears to proceed via the formation of inhomogeneous localised regions of true-vacuum which expand, albeit much slower than the previous case, until they fill the entire space.

\begin{figure}[tp]
    \centering
        \begin{minipage}{0.32\textwidth}
        \includegraphics[width=\textwidth]{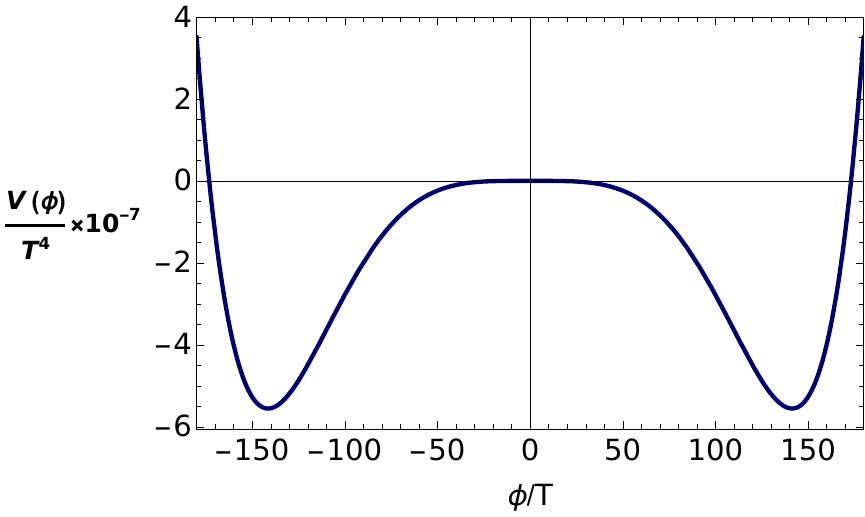}
    \end{minipage}    
    \begin{minipage}{0.32\textwidth}        \vspace{-0.15cm}\includegraphics[width=\textwidth]{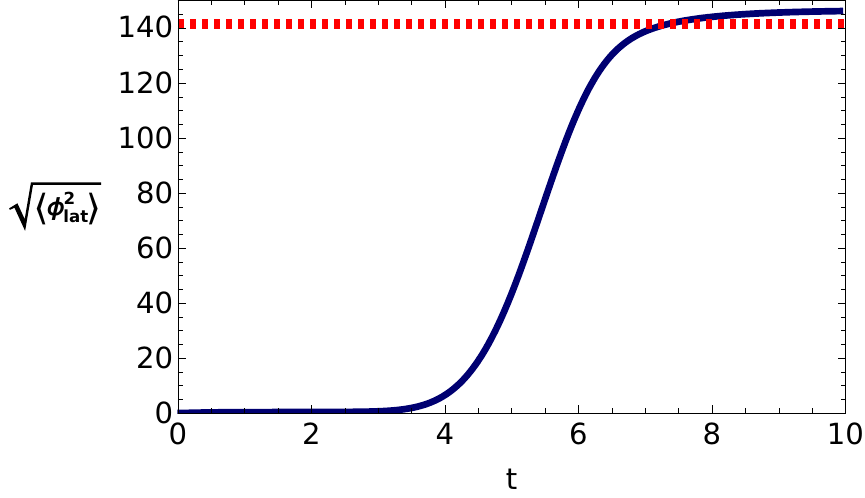}
    \end{minipage}

    \begin{minipage}{0.31\textwidth} 
        \includegraphics[width=\textwidth]{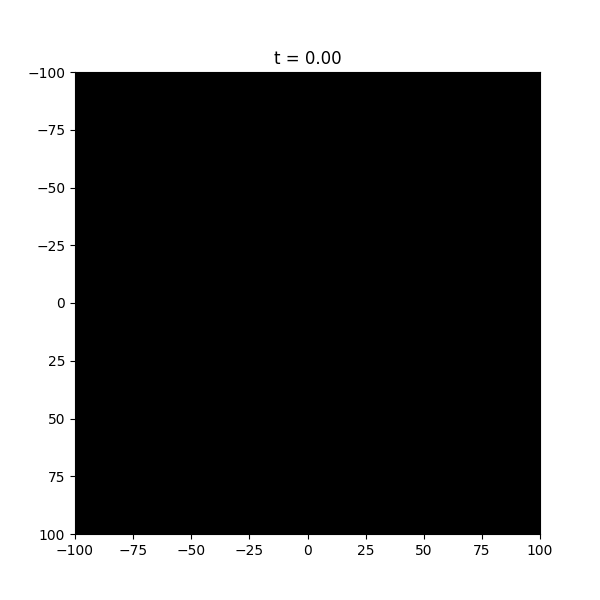}
    \end{minipage}
    \begin{minipage}{0.31\textwidth}
        \includegraphics[width=\textwidth]{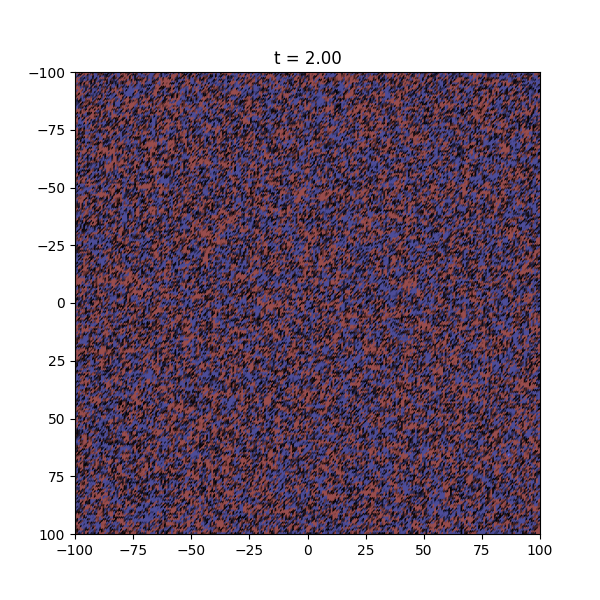}
    \end{minipage}
    \begin{minipage}{0.31\textwidth}
        \includegraphics[width=\textwidth]{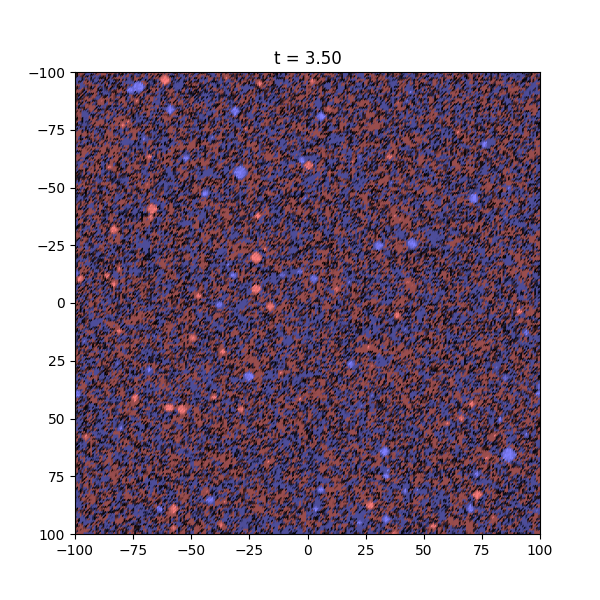}
    \end{minipage}
    
    \begin{minipage}{0.31\textwidth} 
        \centering
        \includegraphics[width=\textwidth]{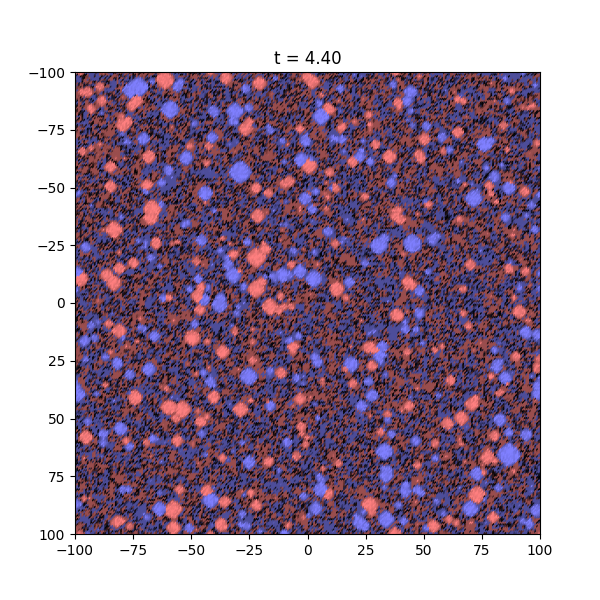}
    \end{minipage}
    \begin{minipage}{0.31\textwidth}
        \centering
        \includegraphics[width=\textwidth]{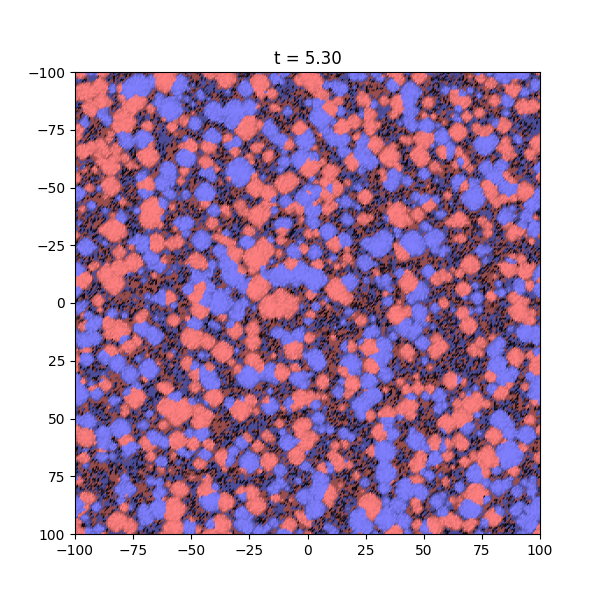}
    \end{minipage}
    \begin{minipage}{0.31\textwidth}
        \centering
        \includegraphics[width=\textwidth]{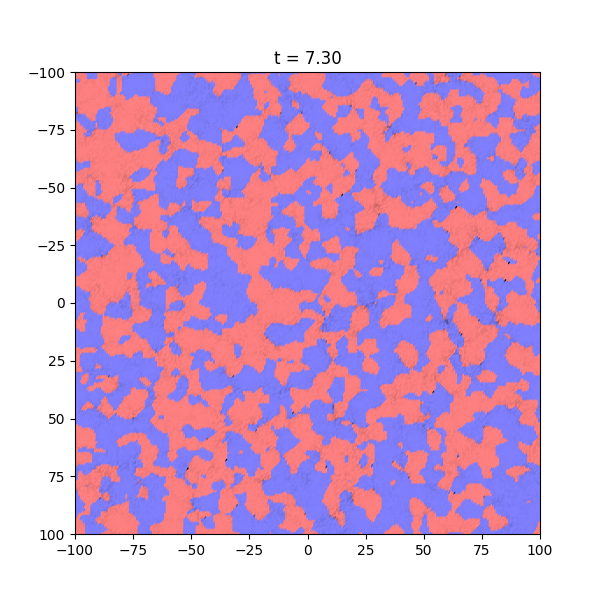}
    \end{minipage}    
    \caption{Benchmark C: \textbf{(Top)} Plot of the potential shape (left) and evolution of the average field value on the lattice, $\sqrt{\langle \phi^2 \rangle }$ (right).
    \textbf{(Below)} Result of the lattice simulation in one $z$-slice at different time steps. The simulation results indicate a phase transition in a first-order-like manner albeit with a very short time scale and a much larger density of bubbles compared to Figures~\ref{fig:benchA} and~\ref{fig:benchB}.}
    \label{fig:benchC}
\end{figure}

Figure~\ref{fig:benchC}, associated with benchmark point C, instead predicts a different time scale for the phase transition compared to the previous scenarios. 
Owing to the exponentially larger predicted nucleation rate of bubbles, we observe a significantly faster total simulation time comparatively. 
This also implies a significantly larger density of true-vacuum bubbles which is clearly observed. 
Nevertheless, there is a clear distinction between pockets of true- and false-vacuum regions, once again indicating that the phase transition nature should be well described by bubble nucleation and expansion. 
For smaller values of $\lambda$ compared to benchmarks A-C, similar behaviour is expected, albeit with exponentially suppressed nucleation rates, see Eq.~\eqref{Eq:S3overT_simulation}. This significantly increases the required simulation time as well as the lattice volume requirements, if a critical density of bubbles is desired within the simulation.

\begin{figure}[!t]
    \centering
        \begin{minipage}{0.32\textwidth}
        \includegraphics[width=\textwidth]{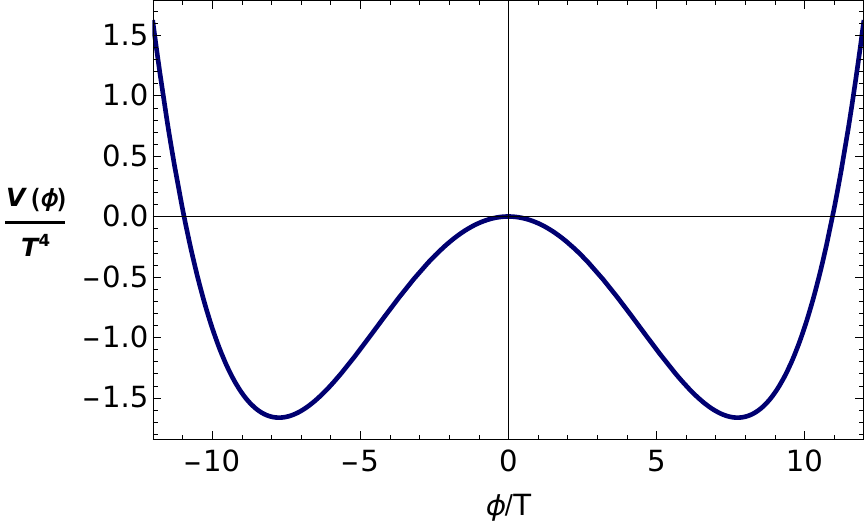}
    \end{minipage}
    \begin{minipage}{0.32\textwidth}        \vspace{-0.15cm}\includegraphics[width=\textwidth]{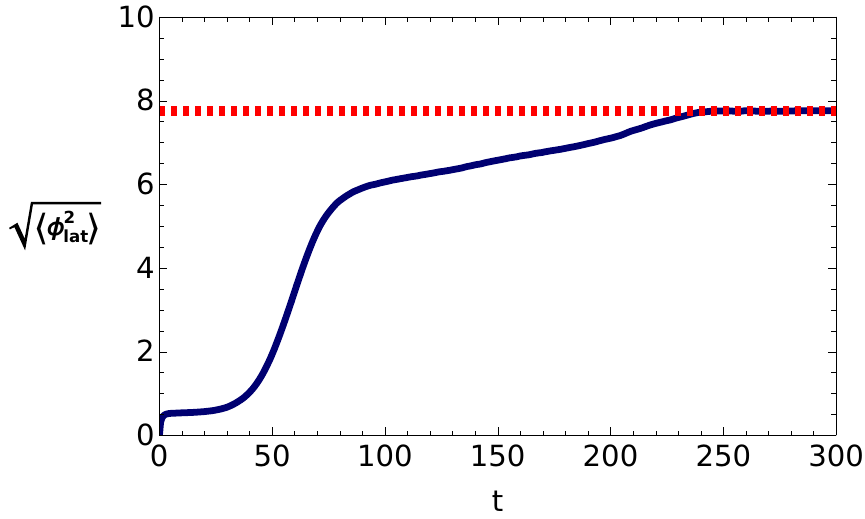}
    \end{minipage}
    
    \begin{minipage}{0.32\textwidth} 
        \centering
        \includegraphics[width=\textwidth]{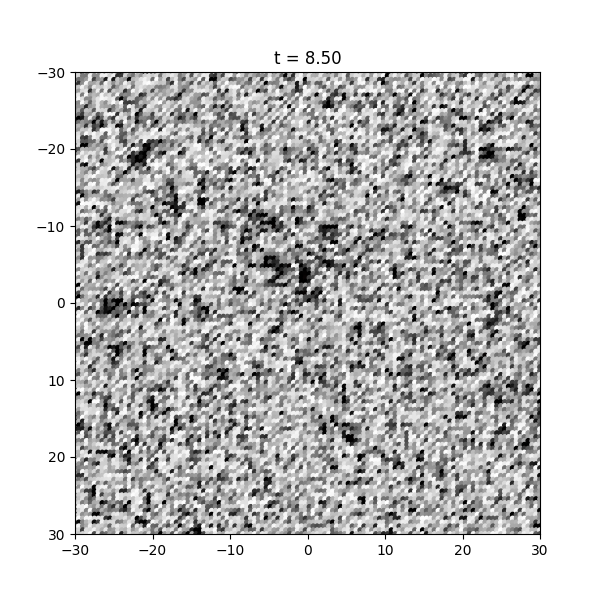}
    \end{minipage}
    \begin{minipage}{0.32\textwidth}
        \centering
        \includegraphics[width=\textwidth]{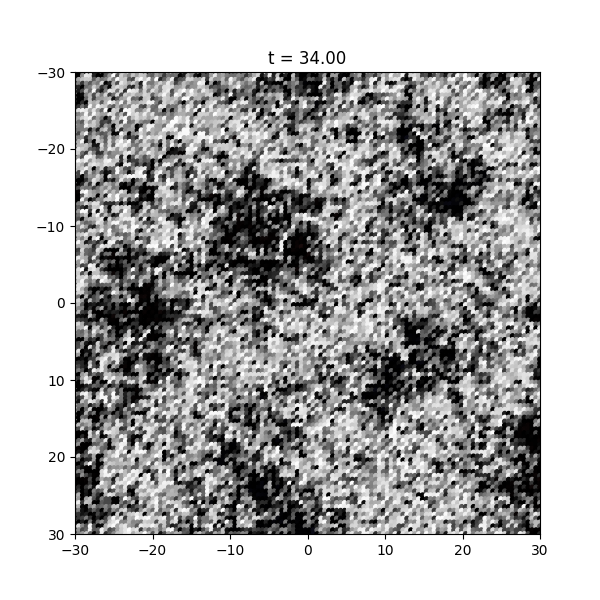}
    \end{minipage}
    \begin{minipage}{0.32\textwidth}
        \centering
        \includegraphics[width=\textwidth]{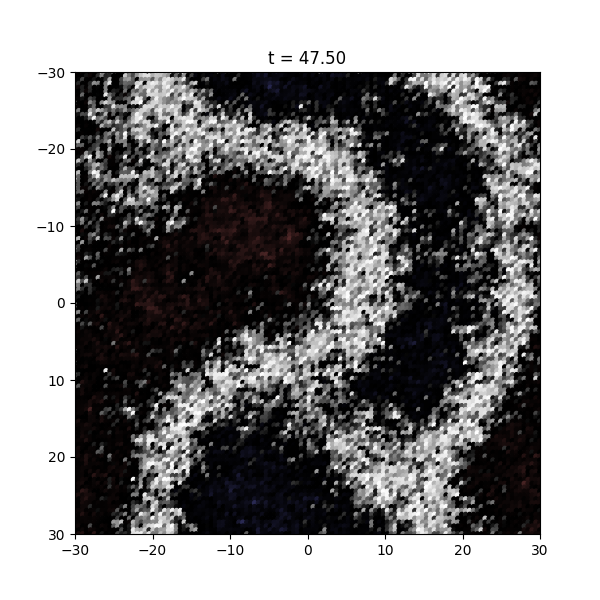}
    \end{minipage}

    \begin{minipage}{0.32\textwidth} 
        \centering
        \includegraphics[width=\textwidth]{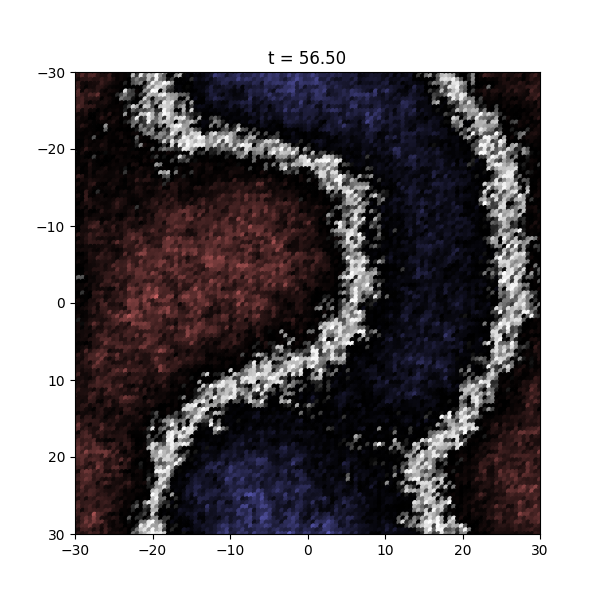}
    \end{minipage}
    \begin{minipage}{0.32\textwidth}
        \centering
        \includegraphics[width=\textwidth]{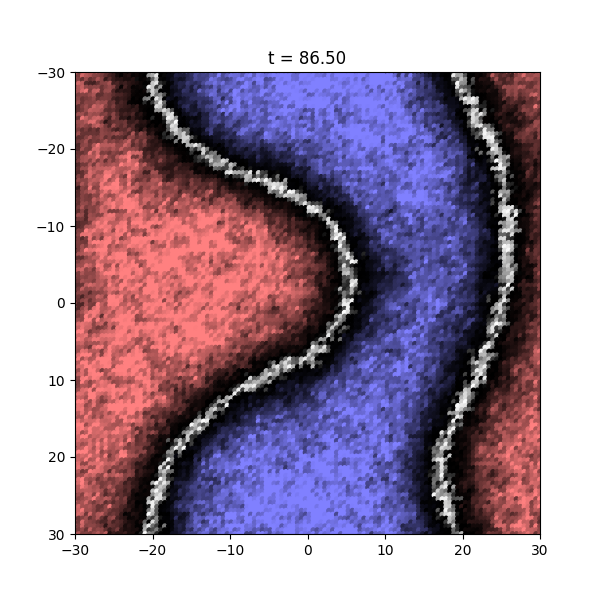}
    \end{minipage}
    \begin{minipage}{0.32\textwidth}
        \centering
        \includegraphics[width=\textwidth]{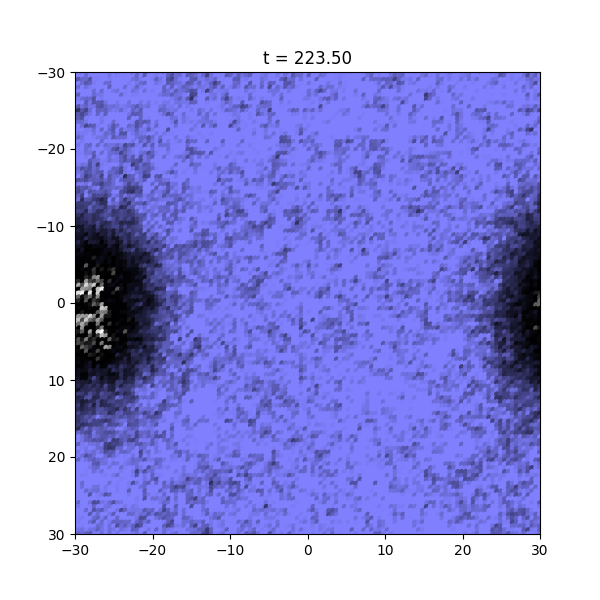}
    \end{minipage}
    \caption{Benchmark M: \textbf{(Top)} Plot of the potential shape (left) and evolution of the average field value on the lattice, $\sqrt{\langle \phi^2 \rangle }$ (right).
    \textbf{(Below)} Snapshots of the simulation of a simple, barrier-less Mexican hat potential with initial condition $\phi=0$. Due to the lack of a potential barrier, we choose here a colour scheme where points go from white (at the origin) to black (around $|\phi| \sim T$) to red or blue near the (degenerate) minima. Compared to the above cases where even a small potential barrier exists, we observe a distinctly different phase transition; that of a homogenous-like roll down of the scalar field, as is expected for such a simulation with no barrier.}
    \label{fig:benchM}
\end{figure}

As all of the benchmark points A-C indicate a phase transition via bubble formation rather than phase-mixing, we simulate benchmark point M and obtain Figure~\ref{fig:benchM}. We choose this benchmark to see the dynamics of a scalar field if it were \textit{not} trapped around the origin by a potential barrier of any size, i.e. the curvature at the origin is negative. 
Instead of well-defined and distinct regions of true- and false-vacuum, we instead observe a smooth, homogeneous evolution of the scalar field as it rolls down to the global minimum. 
Nothing akin to bubble nucleation and expansion is observed.
Although assuming that initially such a field is at the origin is unphysical in a cosmological setting, this simulation demonstrates how significantly 
even a small barrier can impact the qualitative nature of the phase transition. 

As shown in our simulations, $\phi_b < T$ does not ruin the process of bubble nucleation as long as there exists a potential barrier.
We can understand this conclusion in terms of the discussion made in Section.\,\ref{Sec:mot};
the instability that the scalar fluctuation feels is suppressed due to the meta-stable minimum at the origin even though the size of the potential barrier is small compared to the temperature.
Thus, we conclude that time scale for bubble nucleation is much shorter than the time scale of phase-mixing for the case when supercooling is terminated. 
In order to have the time scale of the tachyonic instability to be short enough, the results of benchmark M indicate that a much larger negative curvature around the origin would be required.

\FloatBarrier

\section{Phase transition properties}
\label{sec:SUSYpot}
\subsection{A benchmark model}
In this section, we consider a more motivated toy model arising from the superpotential
\bal
    W = \sqrt{2} \lambda \, \hat \Phi \hat{\bar X} \hat X ,
\label{Eq:W}
\eal
where $\hat \Phi$, $\hat X$ and $\hat{\bar X}$ are chiral supermultiplets.
The factor of $\sqrt{2}$ simplifies the equations below.
We assume that $\hat \Phi$, $\hat X$ and $\hat{\bar X}$ are gauge singlets for simplicity.
The resulting scalar potential of $W$ is,
\bal
V_{\rm SUSY} = 2 \lambda^2 \left[\left| \Phi \right|^2 \left( \left| X \right|^2 + \left| \bar{X} \right|^2 \right) + \left| X \right|^2 \left| \bar{X} \right|^2\right],
\eal
where we denote the complex scalar components of each supermultiplet as $\Phi$, $X$ and ${\bar X}$.
The possible soft SUSY-breaking potential is
\bal
V_{\rm \cancel{\rm SUSY}} = M_{S}\left( \lambda \, A \bar{X} X \Phi + \text{h.c.}\right) + M_S^2 \left(g_\Phi \left| \Phi \right|^2 + g_X \left| X \right|^2 + g_{\bar{X}} \left| \bar{X} \right|^2 \right),
\eal
and for simplicity we take $g_X=g_{\bar X}$ and $A=0$.

Assuming $g_\Phi \ll g_X$ at some messenger scale, there exists a scale at which $g_\Phi=0$.
We denote this scale as $\mu_*$ and evaluate the effective potential at this scale.
The radial mode of $\Phi$ has a nonzero vacuum expectation value (vev) around $\mu_*$ induced from the Colemann-Weinberg potential
\bal
    V_{\rm CW}(\phi) &= \frac{1}{64\pi^2}\Bigg(4\left(g_X M_S^2 + 
    \lambda^2 \phi^2\right)^2 \left( \log\left[\frac{g_X M_S^2 + 
    \lambda^2 \phi^2}{\mu_*^2}\right] - \frac{3}{2}\right) 
    \nonumber \\
    & \qquad \qquad \quad
    - 4\left(\lambda^2 \phi^2\right)^2 \left( \log\left[\frac{\lambda^2 \phi^2}{\mu_*^2}\right] - \frac{3}{2}\right)      \Bigg) 
\label{Eq:VCW}
\eal
where $\phi$ is the radial mode of $\Phi$.
The quadratic term 
\begin{equation}
    \frac{\lambda^2 g_X M_S^2}{8\pi^2} \left( \log\left[\frac{g_X M_S^2 + \lambda^2 \phi^2}{\mu_*^2}\right]-\frac{3}{2}\right) \phi^2
\end{equation} 
becomes negative around the origin when $g_X M_S^2 \ll \mu^2$.
Therefore, assuming that $g_X M_S^2 \ll \mu_*^2$, the effective potential is destabilized at the origin, and a vacuum expectation value (VEV) for $\phi$ is generated.
At large $\phi$, the quartic term $\sim \lambda^4 \phi^4 \log[1+(g_X M_S^2/\lambda^2\phi^2)]$, is negligible compared to the quadratic term, so we obtain
\bal
\label{Eq:v_phi}
v_\phi= \langle \phi \rangle \simeq \frac{ e^{1/4}}{\lambda} \mu_*,
\eal
by minimizing the quadratic term and neglecting $O(M_S^2/\mu_*^2)$ corrections.
The vacuum energy density difference between $\phi=0$ and $\phi=v_\phi$ is given by
\bal
\label{Eq:DeltaV}
\Delta V \simeq \frac{e^{1/2}}{8\pi^2} g_X M_S^2 \mu_*^2.
\eal
Therefore, the effective potential of $\phi$ is expected to be very flat, i.e. 
\bal
\label{Eq:flat_potential}
\frac{\Delta V}{ v_\phi^4}
\simeq \frac{\lambda}{8\pi^2 e^{1/2}} \frac{ g_X M_S^2}{\mu_*^2} \ll 1,
\eal
implying a strongly supercooled first-order phase transition.
Assuming an instantaneous reheating, the reheating temperature can be estimated as
\bal
\label{Eq:T_RH}
T_{\rm RH}
\simeq
1.6 \, \PeV
\left( \frac{100}{g_*} \right)^{1/4}
\left( \frac{g_X^{1/2} M_S}{10\,\TeV} \right)^{1/2}
\left( \frac{\mu_*}{10^{10}\,\GeV} \right)^{1/2}.
\eal

\subsection{Finite-temperature effective potential}
At high temperatures, the $\phi$ potential receives thermal corrections.
Assuming that all components of the supermultiplets are thermalised,
the finite-temperature correction is given by 
\bal
    \Delta V_T(\phi) =  \frac{T^4}{2\pi^2} \left[ 
 4 J_b\left(\frac{m_{\tilde X}^2(\phi)}{T^2}\right) +  4 J_f \left(\frac{m_{X}^2(\phi)}{T^2}\right) \right],
\eal
with
\bal
    J_{b/f}(y^2) = \pm \int_0^{\infty} dx\, x^2 \log\left( 1 \mp \exp(y^2 + x^2) \right).
\eal
Here, $m_{\tilde X}^2 (\phi)=g_X M_S^2 + 
    \lambda^2 \phi^2$ and $m_{X}^2 (\phi)=\lambda^2 \phi^2$ are respectively $\phi$-dependent squared masses of scalar and fermion components of $\hat X$ (and $\hat {\bar X}$).

We include the daisy diagram resummation by taking temperature corrections for the scalar fields.
For $\phi$, we can take second-order derivative of $V(\phi)$, and obtain
\bal
(\Delta m_\phi^2)_T
=\left. \frac{d^2 \Delta V_T(\phi)}{d\phi^2}  \right|_{\phi=0}
=  \frac{T^4}{2\pi^2} \frac{2\lambda^2}{T^2}
\left( 4 J_b'\left(\frac{g_X M_S^2}{T^2}\right)
+ 4 J_f'(0) \right)
\eal
where $J_{b/f}'(x) = dJ_{b/f}(x)/dx $ with $J_b'(0)=\pi^2/12$ and $J_f'(0)=\pi^2/24$.
We also obtain the mass correction to the Goldstone mode in a similar way and find 
\bal
(\Delta m_a^2)_T=(\Delta m_\phi^2)_T.
\eal
In principle, they should be included in the thermal effective potential as $\log(m_0^2 + \Delta m_\phi^2)$ or $J_b((m_0^2 + \Delta m_\phi^2)/T^2)$.
However, the tree-level mass vanishes in our choice of $\overline{\rm MS}$ scale $\mu=\mu_*$, and $(\Delta m^2_\phi)_T$ does not depend on $\phi$.
Therefore, it has no impact on the phase transition dynamics.
For $X$ and ${\bar X}$, we obtain
\bal
(\Delta m_X^2)_T
=
(\Delta m_{\bar X}^2)_T
=
\frac{T^4}{2\pi^2} \frac{2\lambda^2}{T^2}
\left( 2 J_b'\left(\frac{g_X M_S^2}{T^2}\right)
+2 J_b'(0)
+4 J_f'(0) \right).
\eal

Then, the truncated full-dressed effective potential is given by
\bal
& 
\! V_{\rm CW}^{\rm (TFD)} \!\!\!
= \!\!
\frac{1}{64\pi^2} \!
\left\{  \!
4 \! \left(g_X M_S^2 + 
    \lambda^2 \phi^2+(\Delta m_X^2)_T\right)^{\! 2} \!  
    \left(  \! \log \! \left[\frac{g_X M_S^2 + 
    \lambda^2 \phi^2+ \! (\Delta m_X^2)_T}{\mu^2}\right] \! - \frac{3}{2}\right)  
    \right. \nonumber \\
&\left. \qquad\qquad\qquad
- 4\left(\lambda^2 \phi^2\right)^2 \left( \log\left[\frac{\lambda^2 \phi^2}{\mu^2}\right] - \frac{3}{2}\right)      
\right\}   
\\
& 
\Delta V^{\rm (TFD)}_T
=
\frac{T^4}{2\pi^2}   
\left[ 
4 J_b\left(\frac{g_X M_S^2 + \lambda^2 \phi^2+(\Delta m_X^2)_T }{T^2}\right) 
+ 4 J_f \left(\frac{\lambda^2 \phi^2}{T^2}\right)
\right]
\eal
Therefore, the effect of the resummation can be considered as a (temperature-dependent) shift of $g_X M_S^2$.

The critical temperature $T_c$ is defined as the temperature when local minima are degenerate.
Since the effective potential of $\phi$ is flat (see Eq.\,\eqref{Eq:flat_potential}), $\Delta V_T(v_\phi) \simeq 0$ around the critical temperature.
Therefore, $T_c$ can simply be estimated as
\bal
\label{Eq:T2}
\Delta V = |\Delta V_{T=T_c}(0)|= \frac{\pi^2}{12}T_c^4
\quad
\Rightarrow
\quad
T_c \simeq \frac{1}{\pi}
\left( \frac{3 e^{1/2} g_X M_S^2 }{2 \mu_*^2}
\right)^{\!\! 1/4} \mu_* .
\eal
The curvature
of the finite-temperature effective potential at the origin flips its sign at the temperature $T_2$, which can be obtained by solving 
$\left. V_{\rm CW}''(\phi)+\Delta V_T''(\phi)\right|_{\phi=0} =0$.
With the crude approximation $J_b'(g_X M_S^2/T_2^2) \sim J_b'(0)$, this can be estimated as
\bal
T_2 \sim \frac{g_X^{1/2} M_S}{\sqrt{2}\pi } \log \left(\frac{g_X M_S^2}{\mu_*^2} \right)
\label{eq:T2_anal}
\eal
up to a factor of a few uncertainty but this value should ultimately be found numerically.

\begin{figure}[t]
\centering
\includegraphics[width=0.48\textwidth]{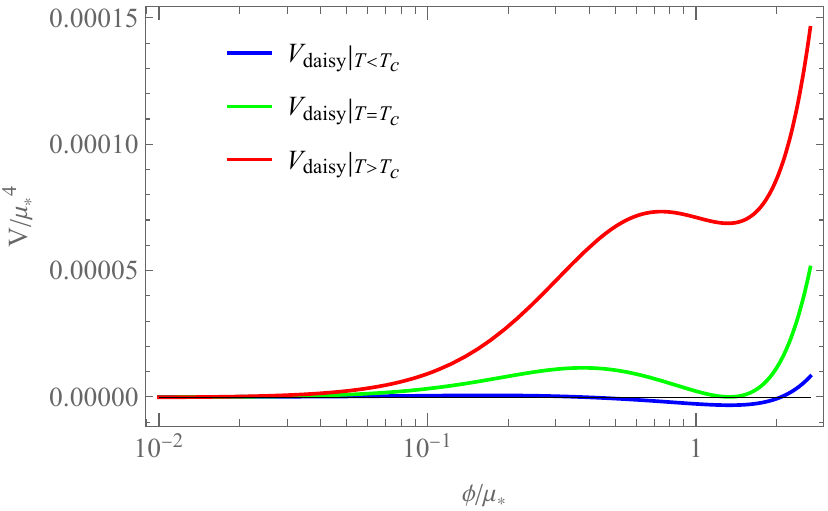}
\caption{Plot of the potential shape around the critical temperature, $T_c$, including contributions from daisy resummation. As $T_c \ll \mu_*$, the second minima forms hierarchically far away from the other two dimensionful scales in the theory: $M_S$ and $T_c$. }
\label{Fig:Vdaisy_Tc}
\end{figure}

\begin{figure}[t]
\centering
\includegraphics[width=0.48\textwidth]{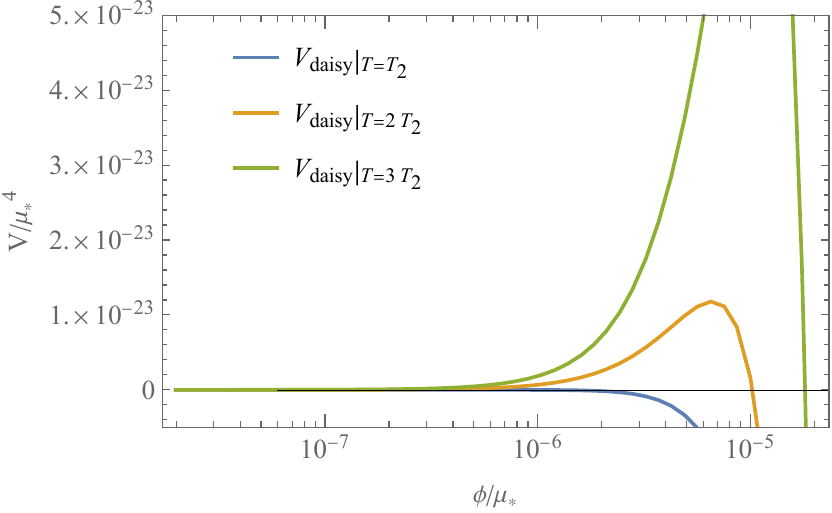}
\includegraphics[width=0.48\textwidth]{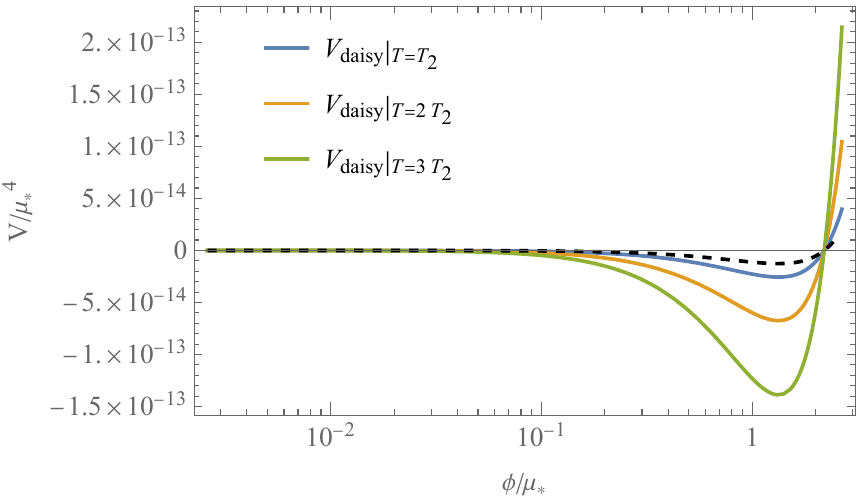}
\caption{Plot of the large variation in the potential shape of the potential around the temperature scale where the curvature flips, $T_2$. Two notable features are present: the barrier very rapidly disappears for small changes in the temperature \textbf{(left)} and the potential remains very flat for a large field distance away from the origin in units of $T_2$ \textbf{(right)}.}
\label{Fig:Vdaisy_T2}
\end{figure}

In Fig.\,\ref{Fig:Vdaisy_Tc} and \ref{Fig:Vdaisy_T2}, we show the temperature dependence of the effective potential, $V_{\rm daisy}$ which includes daisy resummation, around $T_c$ and $T_2$, respectively. This demonstrates the key properties discussed in the previous sections; at the curvature flipping temperature, the barrier rapidly disappears and the potential remains quite flat for large (relative) field values away from the origin.

\subsection{Bubble nucleation rate}

\begin{figure}[t]
\centering
\includegraphics[width=0.48\textwidth]{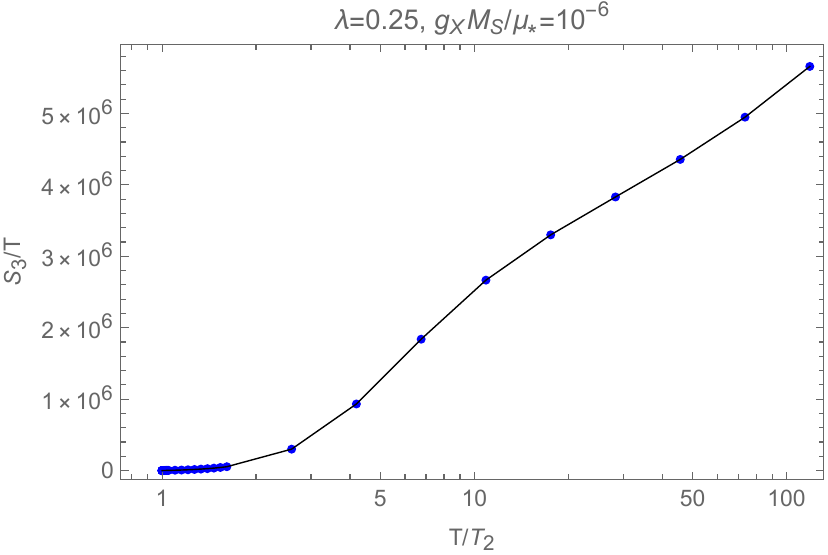}
\includegraphics[width=0.48\textwidth]{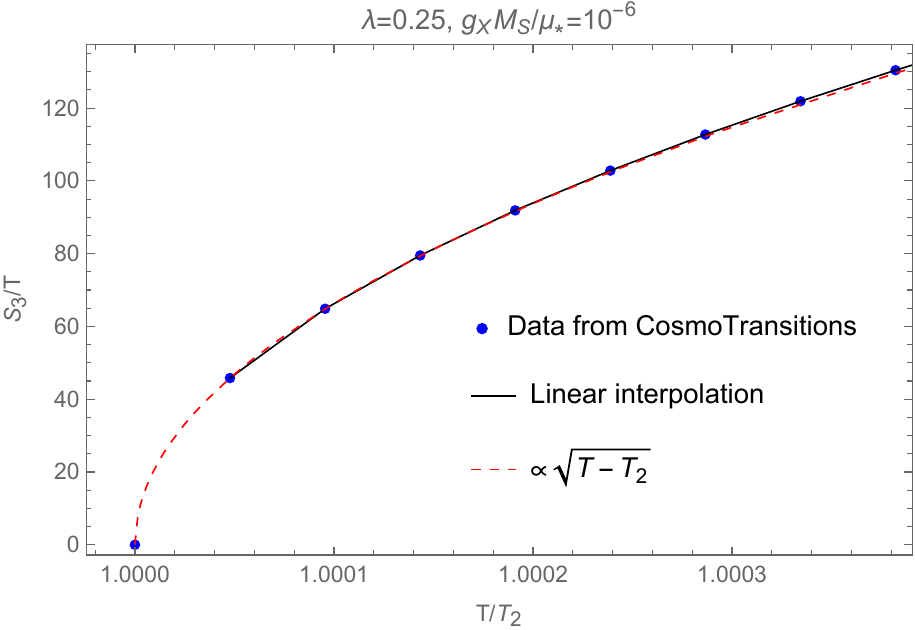}
\includegraphics[width=0.48\textwidth]{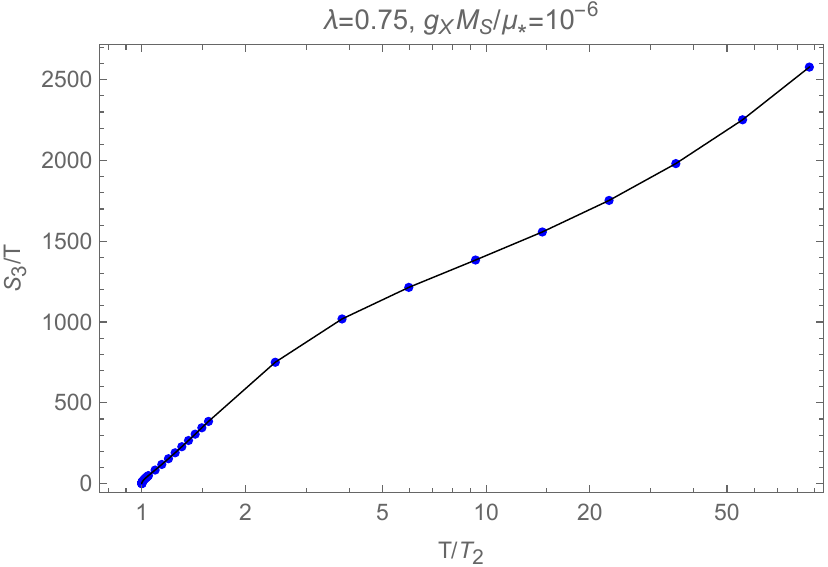}
\includegraphics[width=0.48\textwidth]{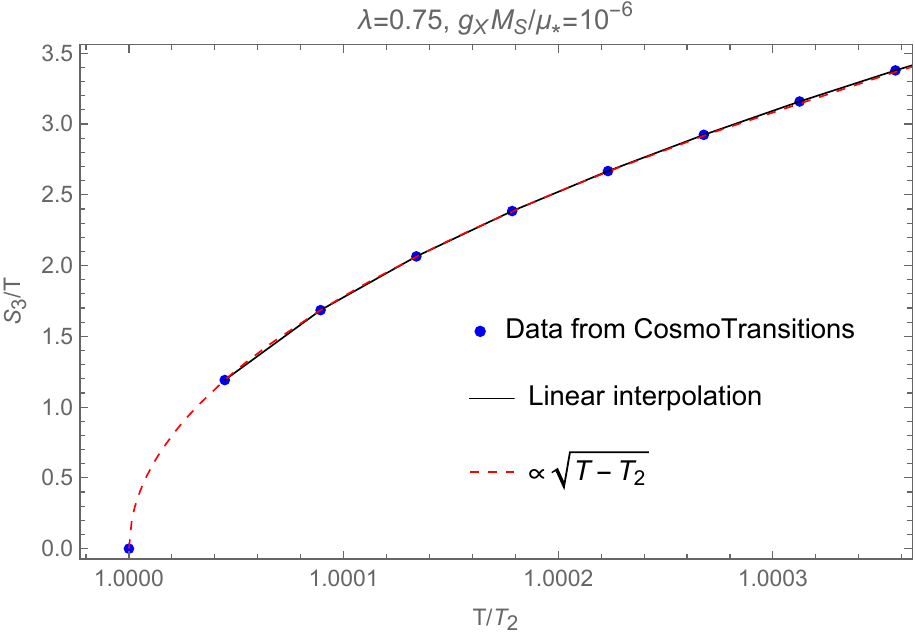}
\caption{Plot of the critical bubble energy, $S_3$, obtained from CosmoTransitions~\cite{Wainwright:2011kj} in two different cases of the example SUSY potential discussed in this section. The variation of $S_3/T$ is given normalised to the curvature-flipping temperature scale $T_2$. The right two plots zoom in for temperatures very close to the $T_2$. Clearly, for temperatures even modestly away from $T_2$, the bubble nucleation rate is highly suppressed and this rapidly changes around $T_2$.}
\label{Fig:S3overT}
\end{figure}

For several sets of parameters, we numerically calculate the bounce action $S_3/T$ at decreasing values of temperature utilising CosmoTransitions~\cite{Wainwright:2011kj} with the results depicted in Fig.\,\ref{Fig:S3overT}.
For the upper (lower) rows, we take $\lambda=0.25$ ($0.75$) with $g_X M_S/\mu_* = 10^{-6}$ and the right-side panel in each case focuses on temperatures very close to $T_2$.

As expected from the origin's curvature flipping at this temperature, $S_3/T$ has a sharp drop around $T_2$.
Around $T_2$, the behaviour of $S_3$ is well approximated by
\bal
\label{Eq:S3overT_analytic}
\left.  \frac{S_3}{T}  \right|_{T\rightarrow T_2} \simeq \frac{k}{\lambda^{3}} \sqrt{\frac{T}{T_2} -1},
\eal
where we numerically find $k\simeq 10^2$.
This can be understood utilising Eq.~\eqref{eq:SUSY_S3T}. 
When $T$ is very close to $T_2$, $\phi_b$ 
is certainly smaller than $T_2$, the barrier is disappearing, and the high-temperature expansion of the potential is valid.
Then, we have
\bal
    V_T(\phi) &\simeq 
    \frac{c_2}{2}  (T^2 -T_2^2)\phi^2 
    - \frac{c_4}{4}  \phi^4 + \cdots ,
\eal
with positive constants $c_2 \sim \lambda^2$ and $c_4 \sim \lambda^4$.
Notice the absence of a cubic term, unlike the SM Higgs case, as the scalar mass has a contribution from the soft breaking term $M_S$: $(g_X M_S^2 + \lambda^2 \phi^2 )^{3/2} \simeq g_X M_S^2 (1 + \frac{3}{2} (\phi^2/g_X M_S^2) \cdots )$.

On the other hand, $S_3/T$ is very large for temperatures even modestly away from $T_2$ (see the left two panels of Fig.\,\ref{Fig:S3overT}).
This implies a long period of supercooling until the temperature becomes comparable to $T_2$, due to of the exponential suppression in nucleation rate, Eq.\,\eqref{Eq:Gamma_3}.
Thus the bubble nucleation temperature, $T_n$, which is defined as $\Gamma_n(T_n) = H^4$, is close to $T_2$ and we can estimate the supercooling parameter $\alpha$ as
\bal
\alpha = \frac{\Delta V}{\rho_{\rm rad}(T_n)}
\sim 
\frac{1}{[\log(g_X M_S^2/\mu_*^2)]^4}
\left(\frac{15 \sqrt{e}}{g_*(T_n)}\right)\left(\frac{\mu_*^2}{g_X M_S^2}\right),
\eal
which is large in the regime $\mu_*^2\gg g_X M_S^2$.

For a detailed study of the phase transition history, the probability that a point is in the false vacuum at time $t$ is given by
\begin{align}
\label{Eq:Pf}
    P_f(t)=\exp\left[ 
    -\int_{t_c}^t dt'\, \Gamma_n(t') \left(e^{H(t-t')}-1 \right)^3 \frac{4\pi}{3 H^3}
    \right],
\end{align}
where $t$ corresponds to cosmological time in FRW coordinates.
As we consider a very strongly supercooled phase transition ($\alpha \gg 1$), we can approximate
\bal
    T (t)=T_n e^{-H (t-t_n)},
\eal
where $T_n$ is the nucleation temperature defined by $\Gamma_n = H^4$ and 
$\Delta V \simeq 3 H^2 M_P^2$ 
with the reduced Planck mass $M_{P}\simeq 2.4\times 10^{18}\,\GeV$.

Eq.\,\eqref{Eq:Pf} is valid only when $t<t_2$ where $t_2$ is the time when $T=T_2$; 
clearly, bubbles are unable to percolate past $t = t_2$.
Since $T_n\simeq T_2$ and $\Gamma_n/H^4 \ll 1$ for $T>T_n$, we can approximate the exponent as
\bal
\label{Eq:I_t}
I(t) \simeq \frac{4\pi}{3} \int_{t_n}^t d(Ht') \left( \frac{\Gamma_n(t')}{H^4} \right)
\, \left(H(t-t')\right)^3.
\eal
From Eq.\,\eqref{Eq:S3overT_analytic}, $\Gamma_n$ can be approximated as
\bal
\frac{\Gamma_n(t')}{H^4}
\simeq
\exp \left[ -\frac{k}{\lambda^3}\left(
\sqrt{\frac{T_n}{T_2}e^{-H (t'-t_n)}-1}
-\sqrt{\frac{T_n}{T_2}-1}
\right)
\right]
\eal
for $H(t'-t_n) \ll 1$,
where we assume that the temperature dependence of the prefactor in Eq.\,\eqref{Eq:Gamma_3} is not as strong as $\exp(- \frac{k}{\lambda^3}\sqrt{T/T_2-1})$ around $T_2$.
On the other hand, $(T_n/T_2-1)$ depends on the symmetry-breaking scale because the dimensionful parameter $H$ comes in to estimate $T_n$.
Again, using Eq.\,\eqref{Eq:S3overT_analytic}, $T_n/T_2-1$ can be written as
\bal
\frac{T_n}{T_2}-1
&=
\frac{\lambda^6}{k^2}
\left[
\log
\left(\frac{9M_{\rm Pl}^4 T_2^4}{64\pi^2 (\Delta V)^2}
\right)
\right]^2
\\
&
\sim
\frac{\lambda^6}{k^2}
\left(O(10^2)
+4\log\frac{T_2}{10\,\TeV}
-2\log\frac{\Delta V}{\PeV^4}
\right)^2,
\eal
assuming $\Gamma_n \sim T^4 \exp(-S_3/T)$.
Thus, we take $T_n/T_2-1 \simeq \lambda^6$, and numerically evaluate $I(t_2)$ by using Eq.\,\eqref{Eq:I_t}.
As a result, we find that 
the percolation occurs before $t_2$, i.e. $I(t_2) > 0.34$, if $\lambda > 0.05$. 
Let us comment briefly about the implications for $\lambda < 0.05$: although such cases do not allow for sufficient percolation by $T = T_2$, at this temperature one should still expect some population of true-vacuum bubbles while the remaining fraction of space is precariously balanced at the (now unstable) origin. 
Depending on the time-scale for the tachyonic instability to run away, i.e. something similar to Eq.~\eqref{eq:sec2_tachyonic}, the nucleated bubbles may still have sufficient time to expand (potentially to much larger radii) and collide.
It is therefore not immediately clear what the fate would be for a phase transition in such a regime.
However, as calculated below, such parameter space predicts relatively large values of the rapidity parameter $\beta$, implying suppressed gravitational wave signals.

\begin{figure}
    \centering
    \includegraphics[width=0.5\linewidth]{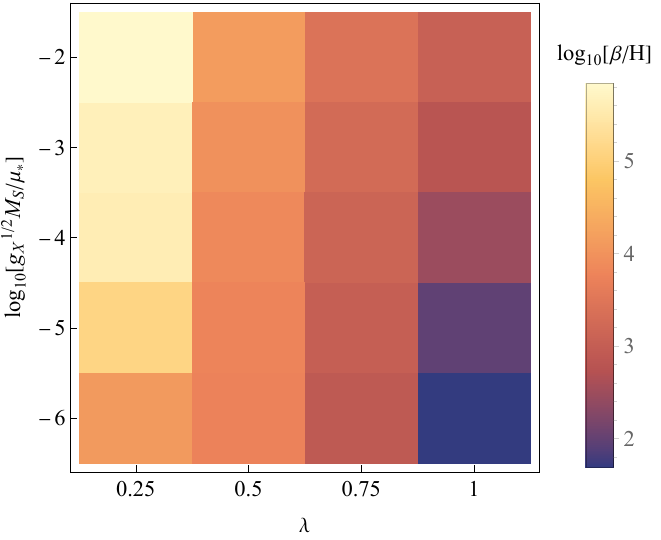}
    \caption{Numerical evaluation of $\beta/H$ as a function of the two relevant parameters of the example model discussed in this section: $\lambda$ and $g_X M_S^2$. A large variation in the time scale of the phase transition is possible and can be as low as $10^2$ for a modestly strong coupling.}
    \label{Fig:beta}
\end{figure}

Although the qualitative picture of the first-order phase transition remains unchanged for $\lambda>0.05$, the very 
small period of cosmological time between $T_n$ and $T_p$ can affect physical observables such as the stochastic gravitational waves produced by the first-order phase transition.
The rapidity parameter $\beta$ is defined by
\bal
\frac{\beta}{H}
=
\left. T\frac{d(S_3/T)}{dT}\right|_{T=T_n}.
\eal
As discussed above, $\lambda$ governs how close $T_n$ is to the temperature of curvature flipping 
at $T_2$, so $\beta/H$ is sensitive to $\lambda$ (and $g_X^{1/2} M_S/\mu_*$).
After numerically evaluating $\beta/H$ in Fig.\,\ref{Fig:beta} for the SUSY potential considered, $\beta/H$ can be as small as $10^2$ for a strong coupling ($\lambda \sim 1$) and a large hierarchy $g_X/\mu_* \sim 10^{-6}$ or as large as $10^6$ for a weaker coupling ($\lambda \sim 0.25$) and a large hierarchy $g_X/\mu_* \sim 10^{-2}$.
In the figure, we also fix $g_X^{1/2}M_S = 10^{4}\,\GeV$ to evaluate $H$ correctly although the final result is only logarithmically sensitive to this choice.

\subsection{Gravitational waves}
Here we provide a brief estimate of the gravitational wave spectrum expected for the predicted bubble-nucleated phase transition. 
An important factor to estimate the gravitational wave spectrum is whether the bubble wall runs away or not.
The maximal pressure on the bubble wall arising from the mass difference of the components of $\hat X$ and $\hat{\bar X}$ on either side of the bubble wall can be estimated by\,\cite{Bodeker:2009qy} 
\bal
{\cal P}_{\rm LO}^{\rm max} = \frac{1}{48}\sum_i g_i n_i \Delta M_i^2 T^2,
\eal
where $g_i$ is the degrees of freedom for the particle $i$ and $n_i=1\,(2)$ when $i$ is fermion (scalar).
$\Delta M_i^2$ is the squared mass difference between the inside and outside of a bubble.
Assuming $g_X^{1/2} M_S \ll \mu_*$, we can estimate $\Delta M_i^2 \simeq \lambda^2 v_\phi^2 \simeq e^{1/2} \mu_*^2$ by using Eq\,\eqref{Eq:v_phi}.
Since $T_n \simeq T_2$, we obtain
\bal
{\cal P}_{\rm LO}^{\rm max} \simeq \frac{e^{1/2}}{16\pi^2} \left[\log\left( \frac{g_X M_S^2}{\mu_*^2} \right) \right]^2 g_X M_S^2\,\mu_*^2
\eal
by using Eq.~\eqref{eq:T2_anal}.
Assuming $g_X M_S^2/\mu_*^2 \gg 1$, ${\cal P}_{\rm LO}^{\rm max}$ is greater than $\Delta V$ in Eq.\,\eqref{Eq:DeltaV}, and thus we conclude that the bubble wall does not run away.

As the bubble wall has a finite velocity, where the pressure coming from the vacuum energy difference is equilibrated with the friction ${\cal P}(v_w)$, the fluid receives work from $\Delta V$, generating a sound-wave profile around the wall\,\cite{Kamionkowski:1993fg, Espinosa:2010hh}. 
The energy budget can be approximately given as totally dominated by the sound-wave contribution and in such a case, the gravitational wave spectrum can be estimated as\,\cite{Hindmarsh:2017gnf}
\bal
\label{Eq:GW_final}
\frac{d \Omega_{\rm GW} h^2}{d\ln(f)}
=
2.061\, F_{\rm GW,0}
h^2 \Gamma^2
\bar U_f^4
(H R_*)
\tilde \Omega_{\rm GW}
S_{\rm sw}(f/f_{p,0}),
\eal
where $F_{\rm GW,0} = 3.5\times 10^{-5} (100/g_*)^{1/3}$ is a red-shift factor of the radiation until today with $h=0.67$\,\cite{Planck:2018vyg}, $\Gamma=1+\omega \simeq 4/3$ is the adiabatic index, $\bar U_f$ is the RMS fluid velocity which we approximate $\sqrt{3/4}$ in the large $\alpha$ limit, $R_*\simeq (8\pi)^{1/3}/\beta$ is the averaged bubble radius, and $\Omega_{\rm GW}\sim 10^{-2}$ is a dimensionless efficiency factor estimated by numerical simulations\,\cite{Hindmarsh:2017gnf, Cutting:2019zws}.
The spectral shape is given by the function $S_{\rm sw}(f/f_{p,0})$
\bal
S_{\rm sw}(x) = x^3 \left( \frac{7}{4+3x^2} \right)^{7/2},
\eal
with peak frequency given by
\bal
f_{p,0} = 8.9\,{\rm Hz}
\left( \frac{(8\pi)^{1/3}\, 10^{-2}}{H R_*} \right)
\left( \frac{z_p}{10}\right)
\left( \frac{T_{\rm RH}}{10^6\,\GeV} \right)
\left( \frac{g_*}{100} \right)^{1/6}.
\eal
Here, $z_p$ parametrizes the actual peak frequency from numerical simulations, which we fix to $z_p\simeq 10$ by following Ref.\,\cite{Hindmarsh:2017gnf}.
As discussed in Ref.\,\cite{Ellis:2018mja}, for the case of a small $H R_*/\bar U_f<1$, Eq.\,\eqref{Eq:GW_final} may lead to an overestimation due to an additional time-shortening effect coming from turbulence, so we estimate the theoretical uncertainty by a factor between $1$ and $H R_*/\bar U_f$.

\begin{figure}
    \centering
    \includegraphics[width=0.5\linewidth]{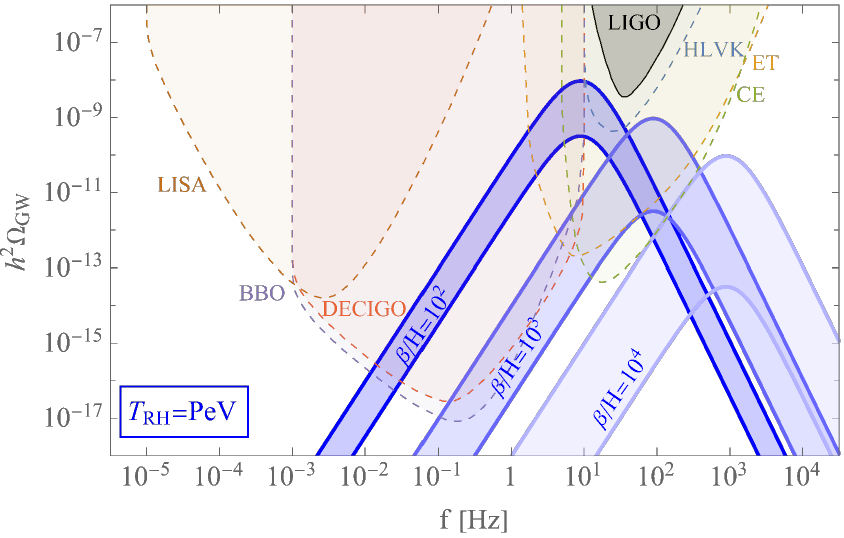}
    \caption{Gravitational wave spectrum and its sensitivity on future gravitational wave detectors for $g_X^{1/2} M_S = 10^4\,\GeV$, and $\mu_* \sim 10^{10}\,\GeV$ leading to $T_{\rm RH} \sim \PeV$ (see Eq.\,\eqref{Eq:T_RH}).
    Here, $\beta/H = 10^2$, $10^3$ and $10^4$ are taken based on Fig.\,\ref{Fig:beta} while the bands represent the uncertainty due to the additional time-shortening effect\,\cite{Ellis:2018mja}.
    Sensitivity curves of LISA, BBO, DECIGO, CE, ET, LIGO and HLVK are taken from the power-law-integrated sensitivity curves in Ref.\,\cite{Schmitz:2020syl}.}
    \label{Fig:GW_sensitivity}
\end{figure}

In Fig.\,\ref{Fig:GW_sensitivity}, we show the resulting gravitational wave spectrum and its sensitivity on future gravitational wave detectors for $g_X^{1/2} M_S = 10^4\,\GeV$ (which should be understood as a soft SUSY breaking parameter), and $\mu_* \sim 10^{10}\,\GeV$ leading to $T_{\rm RH} \sim \PeV$ (recall Eq.\,\eqref{Eq:T_RH}).
As shown in Fig.\,\ref{Fig:beta}, $\beta/H$ ranges between $10^2$ to $10^4$ for $g_X^{1/2} M_S/\mu_* = 10^{-6}$ depending on the coupling $\lambda$, so we depict three cases corresponding to $\beta/H = 10^2$, $10^3$ and $10^4$ while the thickness of each of the bands represent the uncertainty due to the additional time-shortening effect as discussed above.
Sensitivity curves of LISA, BBO, DECIGO, CE, ET, LIGO and HLVK are taken from the power-law-integrated sensitivity curves in Ref.\,\cite{Schmitz:2020syl}.

\section{Conclusion}
\label{sec:conclusion}

We have investigated the fate of the phase transition in the class of supercooled models which exhibit terminated supercooling. 
This termination is the result of a dimensionful mass scale in the effective potential of the phase transitioning scalar field, such that when the temperature of the early universe approaches this mass scale, the curvature of the origin becomes tachyonic. 

As the potential barrier at this temperature scale is quickly disappearing, addressing the question of the fate of such a phase transition is best done with the aid of numerical simulations.
We have established a series of lattice simulations of a scalar field coupled to a thermal bath, with a potential inspired by the supercooled models under study, with: (i) potential barriers smaller than the temperature scale, and (ii) relatively small curvature at and around the origin, predicted by supercooling.
The results of our lattice simulations strongly imply that such a phase transition should proceed via the nucleation and expansion of critical bubbles as opposed to a `phase-mixing'-like transition which can occur when no barrier is present around the origin. 

As we establish that terminated supercooling should generically proceed via bubble expansion, we present a realistic toy model of such a transition and demonstrate that the majority of the motivated parameter space should proceed via bubble formation, i.e. the phase transition is fast enough to occur before the curvature changes sign.
Additionally we estimate the gravitational wave signals that one expects to observe. 
Interestingly we find that the bubble wall velocity is finite due to the small potential difference, and the phase transition rapidity can be sufficiently slow in order for observable signals to appear in future gravitational wave experiments at frequencies above the Hz range. 
The prospects for possible gravitational wave signatures of such models strongly motivate our goal of understanding the nature of the phase transition in such models. 
These models are indeed plausible from the theoretical side, particularly as such sufficiently flat directions in a scalar potential can be realized naturally. 
They also allow for the dilution of long-lived particles in such models which can become problematic when reconciling them with BBN predictions.

Although we conclude that the phase transition in the $m^2$-type flat potential should proceed via bubble formation, there are still further questions unanswered in this work.
These questions include whether the numerical simulations agree with analytic estimates for the nucleation rate and how the spectrum of bubble radii at the onset of bubble collisions compares to the prediction from the usual $\beta$-parametrization of the nucleation rate.
They are beyond the scope of this paper, and we leave them for future work.

\section*{Acknowledgement}

We thank Wan-Il Park and Ke-Pan Xie for useful comments and Oliver Gould for email correspondence. TPD is supported by KIAS Individual Grants under Grant No. PG084101 at the Korea Institute for Advanced Study and thanks Suro Kim for continued and insightful discussions related to this project. This work was supported by IBS under the project code, IBS-R018-D1. CSS is also supported by the NRF of Korea (NRF-2022R1C1C1011840, NRF-2022R1A4A5030362).
\appendix

\bibliographystyle{JHEP}
{\footnotesize
\bibliography{biblio}}
\end{document}